\documentclass[twocolumn,showpacs,preprintnumbers,amsmath,amssymb,aps,prb]{revtex4}
\usepackage{graphicx}
\begin{document}

\title{
Anisotropic Sliding Dynamics, Peak Effect, and Metastability in Stripe Systems 
} 
\author{
C. J. Olson Reichhardt, C. Reichhardt, and A.R. Bishop} 
\affiliation{
 Theoretical Division,
Los Alamos National Laboratory, Los Alamos, New Mexico 87545 \\} 

\date{\today}
\begin{abstract}
A variety of soft and hard condensed matter systems are known to form 
stripe patterns.
Here we use numerical simulations to
analyze how such stripe states depin and slide when 
interacting with a random substrate and with
driving
in different directions with respect 
to the orientation of the stripes. Depending on the
strength and density of the substrate disorder, we find that 
there can be pronounced anisotropy in the transport
produced by different dynamical flow phases. 
We also find a disorder-induced ``peak effect'' similar to that observed for
superconducting vortex systems, which 
is marked by a transition from elastic depinning to a state
where the stripe structure fragments or partially disorders at depinning. 
Under the sudden application of a driving force, 
we observe pronounced metastability effects similar to 
those found near the order-disorder transition associated with the
peak effect regime for three-dimensional superconducting vortices. 
The characteristic transient time 
required for the system to reach a steady state 
diverges in the region where the flow changes from elastic to disordered.
We also find that anisotropy of the flow 
persists in the presence of thermal disorder
when thermally-induced particle hopping along the stripes 
dominates.
The thermal effects can wash out the 
effects of the quenched disorder, leading to 
a thermally-induced stripe state.
We map out the dynamical phase diagram for this
system, and discuss how our results 
could be explored in electron liquid crystal systems, 
type-1.5 superconductors, and pattern-forming colloidal
assemblies.   
\end{abstract}
\maketitle

\vskip2pc

\section{Introduction}

Stripe formation occurs in a wide variety of 
soft \cite{Andleman,Boyer,Singer,Pell,Glaser,Colloid,S,Colloid2,Zapperi,Peeters}
and hard 
\cite{J,Stripe,SA,Bishop,Kabanov,Stroud,Kivelson}
condensed matter systems. 
These stripe patterns are often a consequence of
some form of effective competing or multiple 
length scales in the pairwise interactions between the particles 
\cite{Andleman,Singer,Pell,Glaser,Colloid,Zapperi,Peeters,Bishop,Bishop2,Kabanov,Stroud}. 
For soft condensed matter,
pattern formation can occur when the particles experience
intermediate range repulsion and short range attraction,
such as in certain types of colloidal systems 
\cite{Colloid,Zapperi}.
In addition to stripe phases, numerous other patterns can appear
as a function of density, temperature, or particle interaction strength, 
including bubble, clump, and uniform crystalline phases
\cite{Andleman,Singer,Pell,Glaser,Bishop,Peeters,Kivelson,Bishop3}. 
The competing interactions that produce the stripes
may be produced by particles that have both a short range attraction
and a long range repulsion
\cite{Bishop,Bishop2,Bishop3,Peeters}; 
however, systems 
with only repulsive interactions 
can also exhibit stripe phases \cite{Pell,Glaser,S,Colloid2} 
provided that there are at least two length scales
in the interaction potential. 
Typically, as the density increases, the system progresses from a
low density clump phase to an intermediate density stripe phase, and then to
a higher density bubble phase where organized voids appear in the system;
finally, at the highest densities, the particles form a uniform crystal state
\cite{Pell,Colloid2,Zapperi,Bishop3}. 
In two-dimensional (2D) systems 
of finite size,
a stripe phase containing oriented stripes is often observed 
\cite{Pell,Glaser,Bishop3}; however,  
for larger systems, 
the strong degeneracy in the stripe ground state orientation can produce
a labyrinth pattern composed of many different stripe orientations
\cite{Singer,Pell,Colloid2,Martin}.  The presence of any type of
bias produced by
the boundaries, a substrate, or an external drive such as a
shear breaks the symmetry of the stripe ground state and causes 
the stripes to align in a single direction \cite{Martin,Olson,Zapperi}.

In addition to soft matter systems,
there is growing evidence that stripe and 
bubble phases occur in hard condensed matter
systems such as 2D electrons in the quantum Hall regime 
\cite{Kivelson,Lilly,Du} 
and charge ordering in 
high temperature superconductors \cite{J,Stripe,SA,Bishop,Kabanov,Stroud}.
Evidence for stripe phases in two-dimensional electron gas (2DEG) systems 
includes anisotropic transport curves which 
have been interpreted as indicating that
the stripes have a single preferred orientation \cite{Lilly,Du,JE,M}. 
If the stripes take the form of a charge ordered state,
the transport anisotropy implies
that the stripes can slide more easily
when the drive is applied parallel to the stripes than when it is applied
perpendicular to them.
The alignment of the stripes in the 2DEG systems may be due to 
small intrinsic biases that form during sample growth \cite{Co}. 
There have also been recent 2DEG experiments 
that show that dc drives or other external driving can 
dynamically orient the stripes under certain conditions \cite{Gores,Cheong}.
Other recent experiments have shown that 
the stripe direction can be controlled with a 
strain, making it possible to alter
the anisotropy with a strain field \cite{Gabor}. 
Transport experiments 
in 2DEGs have revealed sharp conduction thresholds, a  
series of intricate jumps in the current versus resistance curves, 
pronounced hysteresis, and changes in the conduction noise, 
suggesting that these systems are
undergoing depinning transitions and dynamic changes in the sliding dynamics 
\cite{Cooper,N,Gores}.
Additional evidence for charge ordered states in 2DEGs 
has come from resonance measurements
\cite{Lewis,Zhu}. 
  
Another recently described system where stripe 
patterns occur is in ``type-1.5'' superconductors, predicted to appear in
two-band superconductors such as MgB$_{2}$ \cite{Mosch}.
In a type-II superconductor under a magnetic field, 
the flux in the sample takes the form
of quantized vortices which organize into a uniform 
triangular lattice as a result of 
their repulsive interactions.
In contrast, in type-1.5 superconductors,    
the vortices have both an attractive and a 
repulsive component to their interactions \cite{Mosch,Mosch2},   
which in principle will lead to the formation of clumps and stripes.  
Experiments in the 
two-band superconductor systems have 
revealed evidence for disordered clump-like vortex structures;  
however, strong pinning in 
the samples probably prevents the detection of
ordered patterned structures \cite{Mosch}. 

In the 2D electron systems and the type-1.5 superconductors, 
the interplay between the disorder in the sample
and an external drive 
should produce very rich dynamics
with different types of depinning transitions and sliding 
states.  It should also be possible to 
subject stripe-forming soft matter systems 
to both an external drive and quenched disorder. 
Experiments have already been conducted on the
depinning of purely repulsive colloids interacting
with quenched disorder \cite{Ling2}. 
Similar experiments could be 
performed with colloidal systems 
that have interactions which lead to stripe formation. 
The work we describe here is also relevant to 
systems exhibiting anisotropic sliding friction due to the
formation of stripe-like surface ordering \cite{Hu}. 
To address how stripe-forming systems behave in the presence of both
driving and quenched disorder, 
we simulate a collection of particles driven over randomly placed
attractive pinning sites and interacting 
with a long range repulsion and a short range 
attraction \cite{Olson,Bishop,Bishop3}.
We show that when the stripes have a specific orientation, 
a number of distinct sliding states can occur which have distinct
anisotropic transport signatures, 
including a peak in the anisotropy 
produced when the dynamics is plastic 
for driving transverse to the stripes but elastic for driving along the
stripes.

In our previous work, we have examined the depinning and sliding of disordered
bubble, clump, and stripe phases, and found that for a fixed pinning 
density, the stripe phase has the highest depinning threshold 
\cite{Olson,Bishop}. 
In this case, the stripes did not have a single orientation but instead formed
a disordered labyrinth pattern.  When plastic depinning occurred, it was
possible to induce a dynamical reordering transition into a stripe state
aligned with the driving direction \cite{Olson}. 
We also found that the dynamically induced reorientation 
strongly depends on the strength of the quenched disorder.
Only for sufficiently strong quenched disorder are
there enough plastic distortions to permit the formation of the oriented
stripes \cite{Bishop}.  
When the pinning is weak, the labyrinth structures depin 
elastically without any distortions
and the aligned stripes never form. 

Here we analyze the transition
from elastic to plastic depinning 
and show that for some parameters, the stripe system exhibits 
a peak effect phenomenon similar to 
that observed at
the transition from elastic to plastic 
depinning in vortex matter. 
The vortex peak effect is associated with a sharp increase in the 
depinning force as well as changes in the
transport curves \cite{Peak,Higgins,Ling,P}. 
We show that the peak effect 
in the stripe system can occur for driving in either
direction and that it is possible to have a peak effect 
for one direction of drive but not the other. 
In previous work, we showed that there is a broad 
maximum in the depinning force 
for the stripe phase as function of the strength of the attractive term. 
In this work, we study the peak effect in the stripe phase 
as a function of disorder and temperature, and find that it 
occurs as a sharp, first-order-like transition which is 
similar to the peak effect observed in superconductors.

In this work we explicitly focus on the case where the stripes 
are already in an aligned state rather than in a disordered labyrinth phase. 
This permits us to apply a drive in two well-defined directions, along and
perpendicular to the stripes, and to compare the anisotropic response for
different strengths of quenched disorder.
As noted previously, many of the 2DEG stripe systems appear to contain
oriented stripes.
To our knowledge the depinning and sliding dynamics of 
an oriented stripe system has not previously been numerically studied. 
We find several new types of 
sliding phases that do not appear for
sliding dynamics in isotropic systems such as vortices 
\cite{Koshelev,Balents,Zimanyi,Olson2,Marchetti,Pardo,Brass},
colloids \cite{Ling}, sliding charge density waves
\cite{Du2}, or sliding Wigner crystals \cite{Jensen}.
For example, we find several different types of plastic stripe flow.
In one state, individual structures slide past stationary stripes; 
in another state, the stripe structure remains intact but
a portion of the particles within the stripes are pinned 
while other particles flow past in one-dimensional (1D) channels.
For strong disorder the stripe structure breaks apart and the 
flow is similar to the plastic flow observed in 
isotropic vortex systems \cite{Zimanyi,Olson2,Brass}. We 
also find that the extent to which the stripes reorient is strongly 
sweep rate dependent.  This affects measurements of the depinning
thresholds and 
features in the transport curves.
     
\begin{figure}
\includegraphics[width=3.5in]{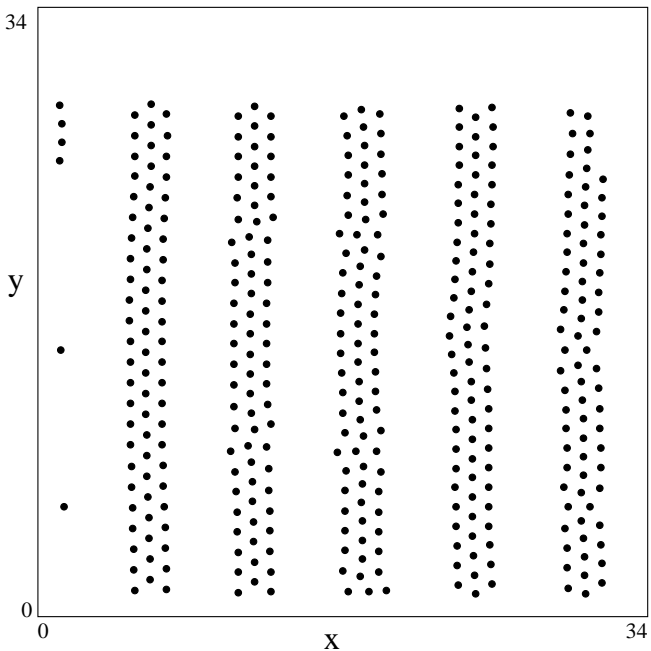}
\caption{
The particle positions (black dots) 
for a system with competing long-range repulsion and 
short-range attraction.  For this density of 
$\rho=0.36$ we obtain a stripe state with the stripes aligned 
along the $y$-direction. 
The external drives are applied either along the stripe direction,
$F^{y}_{D}$, 
or perpendicular to the stripes, $F^{x}_{D}$.  
}
\end{figure}

\section{Simulation}

In Fig.~1 we show a snapshot of our system 
containing stripes oriented in the $y$-direction.
The easy driving direction is along $y$, parallel to the stripes, while
the hard driving direction is along $x$, perpendicular to the stripe pattern.
Our simulation box has periodic boundary conditions 
with sides $L_y=L$ and $L_x=1.097L$.  We consider $N=380$ particles
with a density given by $\rho = N/(L_xL_y)$.
Here we fix $\rho=0.36$.
The particles interact with a long range 
Coulomb repulsion and a short range exponential attraction. 
The resulting interaction potential is 
repulsive at very short ranges due to the Coulomb term, 
attractive at intermediate range, and repulsive at long range. 
The dynamics of the particles are determined by integrating 
the following equation of motion:   
\begin{equation}  
\eta \frac{d {\bf R}_{i}}{dt} = 
-\sum_{j\ne i}^{N_{i}}{\bf \nabla}V(R_{ij}) +  {\bf F}^{P}_{i} +  {\bf F}^{DC}_{i} +  {\bf F}^{T}_{i} .
\end{equation} 
Here ${\bf R}_{i}$ is the position of particle $i$.

The first term on the right hand side of Eq.~(1) is the particle-particle 
interaction potential
\begin{equation} 
V(R_{ij}) =  \frac{1}{R_{ij}} -B\exp(-{\kappa} R_{ij})  
\end{equation} 
with $R_{ij} = |{\bf R}_{i} - {\bf R}_{j}|$, $B=2.0$, and $\kappa=1.0.$ 
To avoid the divergence from the Coulomb term  
at small $R_{ij}$ we place a constant-force cutoff at $R_{ij}<0.1$. 
The Coulomb term does not permit a long range interaction cutoff 
so for computational efficiency we 
employ a Lekner summation method to calculate the long range Coulomb force 
 \cite{Lekner}. 
The second term 
of the interaction potential
is a phenomenological short range attractive interaction.

The second term on the right in Eq.~(1)
is the force from the quenched disorder,  modeled as 
$N_{p}$ non-overlapping 
randomly placed parabolic pinning sites 
with density $\rho_p=N_p/(L_xL_y)$ and with 
\begin{equation}
{\bf F}^{P}_{i} = \sum_{k=1}^{N_p}\left(F_{p}/R_p\right)R_{ik}^{(p)}
\Theta(R_{p}- R_{ik}^{(p)}){\bf \hat{R}}_{ik}^{(p)}. 
\end{equation}
Here, ${\bf R}_k^{(p)}$ is the location of pinning site $k$, $R_p$ is the
pinning radius which is set
to $R_p=0.2$ unless otherwise noted, 
$F_p$ is the maximum force from a pinning site,
$R_{ik}^{(p)}=|{\bf R}_{i}-{\bf R}_k^{(p)}|$, 
${\bf {\hat R}}_{ik}^{(p)} = ({\bf R}_{i} - {\bf R}_{k}^{(p)})/R_{ik}^{(p)}$, 
and $\Theta$ is the
Heaviside step function. 

The force ${\bf F}^{DC}_i$ in Eq.~1 arises from 
an external dc drive applied unidirectionally 
to all the particles in either the $y$ or $x$ direction, 
${\bf F}^{DC}_i = F_{D}^y{\bf {\hat y}}$ or
${\bf F}^{DC}_i = F_{D}^x{\bf {\hat x}}$.
We measure the depinning threshold and 
transport curves for each driving direction by summing over the 
velocities of the particles, 
$\langle V_\alpha\rangle= \sum^{N}_{i} {\bf v}_{i}\cdot {\bf \hat \alpha}$ 
with $\alpha=x,y$.

The final term on the right hand side of Eq.~(1)
represents the forces from randomly 
distributed thermal kicks with the following properties:   
$\langle F^{T}_i(t)\rangle = 0$ 
and $\langle F^{T}_{i}(t)F^{T}_j(t^{\prime})\rangle = 2\eta k_{B}\delta_{ij}(t - t^{\prime})$,   
where $k_{B}$ is the Boltzmann constant.
In previous equilibrium studies of this system, 
we identified the densities at which different clump, stripe, and bubble 
phases occur \cite{ourstuff}.
Here we work at $\rho=0.36$
corresponding to the case of stripes containing 
approximately three particles per row as shown
in Fig.~1.
The initial particle positions were obtained from a very slow
simulated annealing from a high $T$ 
to $T = 0.0$.  The stripes align in the $y$-direction during the anneal,
and the pinning potential is not applied until after the annealing process.  

\section{Pulse Measurements and Metastability for Strong Disorder} 

We first examine the dynamical response when different strengths of
external drive are suddenly applied to the system.
When the external drive is slowly increased from zero, the system passes
through several different dynamical phases.  In contrast, for the sudden
pulse drive, the system can pass directly from a pinned state to a sliding
state, and an ordered moving state may appear that cannot be reached
by slowly increasing the driving force.
For strong pinning, the stripe structure breaks 
up or fragments near depinning, and the anisotropy of the two
driving directions is reduced for a slow ramp of the driving force; 
however, for the pulse measurements, a pronounced anisotropy can be
preserved.

\begin{figure}
\includegraphics[width=3.5in]{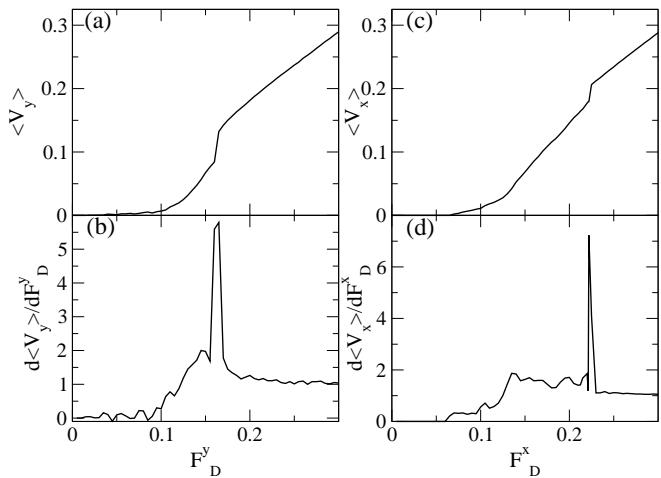}
\caption{
(a) The average steady state velocity 
$\langle V_y\rangle$ vs $F^y_D$ for pulse driving in the $y$-direction.
We use the same parameters as in Fig.~1 with $F_p=0.9$
and $\rho_p=0.38$.
Here the drive is suddenly increased from zero to $F_D$
and the system settles into a steady state after a transient time
$\tau$. 
(b) The corresponding $d\langle V_{y}\rangle/dF^{y}_{D}$, 
with a sharp peak indicating the transition
to a moving stripe phase for $F^{y}_{D} \ge 0.165$. 
(c) $\langle V_{x}\rangle$ vs $F^{x}_{D}$ for the same
system. (d) The corresponding 
$d\langle V_{x}\rangle/dF^{x}_{D}$ 
vs $F^{x}_{D}$ showing a sharp peak at the transition
to a perpendicularly translating stripe state. 
}
\end{figure}

\begin{figure}
\includegraphics[width=3.5in]{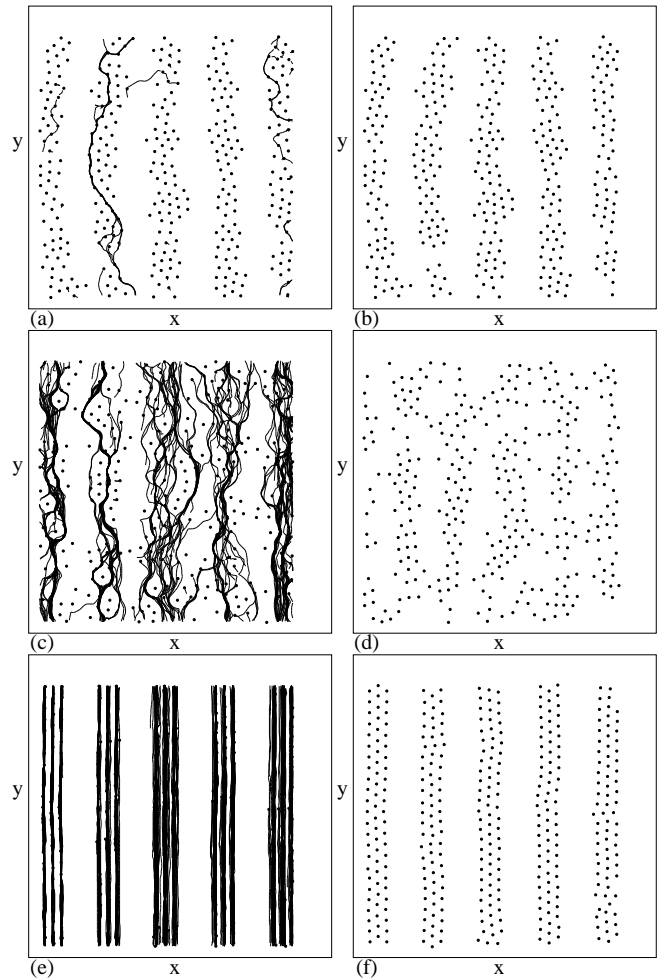}
\caption{ 
The particle positions (black dots) and particle trajectories 
(black lines) for the system in Fig.~2(a,b)
for pulse driving in the $y$-direction. 
The particle trajectories are traced over equal times in panels (a), (c),
and (e).
(a) At $F^{y}_{D} = 0.05$ there is a filamentary flow pattern 
along the stripes. 
(b) A snapshot of only the particle positions from (a) shows that
the stripe structure is partially preserved.
(c) At $F^{y}_{D} = 0.125$, fluctuating plastic flow 
occurs in which channels of moving particles intertwine and mix 
while other particles remain pinned. 
(d) The particle positions only for the system in 
(c) indicate that the stripe structures are completely
disordered. 
(e) At $F^{y}_{D} = 0.2$, above the peak in 
$d\langle V_{y}\rangle /dF^{y}_{D}$ shown in Fig.~2(b), 
all the particles are moving in an ordered stripe phase. 
(f) A plot of only the particle positions from (e) shows
 the ordered stripes.
}
\end{figure}

We conduct a series of pulse drive simulations at $F_{p} = 0.9$
and $\rho_{p} = 0.38$.  For these parameters, 
slow driving ramps would produce a breakup of the stripe structure for
driving in either the $x$ or $y$ direction. 
After we apply the pulse drive, the system typically passes through a
transient state and the velocity relaxes to a steady state
value after a characteristic time $\tau$.
We construct a pulsed-drive velocity-force curve by plotting the
average steady state velocity $\langle V\rangle$ versus the magnitude of
the pulse drive $F_D$.  This is shown in
Fig.~2(a) and Fig.~2(c) for driving along $y$ and $x$, respectively.
For driving in the easy or $y$-direction, the critical force 
$F^{y}_{c}$ is lower than for driving in the       
hard or $x$-direction. 

There is a sharp jump 
to a higher value of $\langle V_{y}\rangle$ 
just above $F_{D}^y = 0.165$ in Fig.~2(a).
For $F_D^y<0.165$, the pinned stripe state undergoes plastic distortion when
it moves, but for $F_D^y\ge 0.165$, the pinned stripe is able to depin
directly into an ordered moving stripe state, resulting in the jump in
mobility.  An example of the ordered motion is shown in 
Fig.~3(e) and Fig.~3(f) for $F_{D}^y = 0.2$.
For $F^y_{D} < 0.165$, the stripe structures are partially destroyed when
plastic flow occurs at depinning.  For weakly plastic flow, 
the moving stripes persist
transiently for a period of time before breaking apart and repinning,
as in Fig.~3(a) and Fig.~3(b) at $F^y_D=0.05$.
In contrast, the stripes are disordered in the strongly fluctuating 
plastic flow phase when a portion of the particles are pinned while other 
particles are mobile, 
as illustrated in Fig.~3(c) 
and Fig.~3(d) for $F_{D} = 0.125$.
The transition from the disordered plastic flow regime to the
ordered moving stripe regime 
appears as a pronounced peak
in $d\langle V_y\rangle/dF^{y}_{D}$ at $F^y_D=0.165$, as shown in Fig.~2(b). 

\begin{figure}
\includegraphics[width=3.5in]{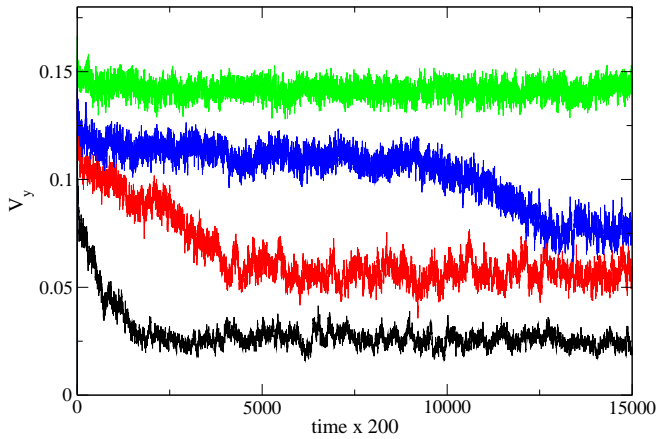}
\caption{
The velocity $V_y$ in the $y$-direction averaged over 200 simulation time
steps vs simulation time for pulse drives
of $F^{y}_{D} = 0.11$, 0.125, 0.15, and $0.17$, from bottom to top. 
In the lower three curves, which fall in the
strongly fluctuating plastic flow regime,
the transition from a higher to a lower
velocity occurs when the ordered stripe structure breaks apart,
such as when the structure seen in Fig.~3(f) 
turns into a fragmented structure of the type shown in Fig.~3(d). 
The upper curve is at a drive above the transition to the moving
ordered stripe regime.
}
\end{figure}

For the strongly fluctuating plastic flow regime
found for $0.1 < F^{y}_{D} < 0.165$, the initial transient motion consists of
an elastically moving stripe state in which all the particles are moving.
To characterize the time $\tau$ required before the system reaches
a steady state after application of a pulse drive,
we analyze the time series of $V_{y}$ such as those plotted in Fig.~4 
for $F^y_{D}  = 0.11$, 0.125, $0.15$, and $0.17$, 
where each point is averaged over 200 simulation time steps.
The system starts with a 
higher value of $V_{y}$ which persists for a time that increases with
increasing $F^y_{D}$ before dropping to the lower steady state value
of $V_{y}$. 
The initial motion associated with the higher value of $V_y$ is a metastable
moving ordered stripe.  After the transient time $\tau$, the stripe breaks
apart and a portion of the particles become pinned, producing the drop to
the lower value of $V_y$.
The particle positions shown in Fig.~3(d) at $F_{D}^y = 0.125$ 
are illustrated at a point in time 
after the transient ordered stripe state broke apart.

For $F^y_{D} \ge 0.165$, the system remains 
in the ordered moving stripe state within 
the entire simulation time window, which includes
simulations ten times longer than shown in Fig.~4.  
For $F^y_D<0.165$, the transient time $\tau$ during which the metastable
ordered moving stripe exists increases with 
increasing $F^y_{D}$. 
It is possible that 
after extremely long times, even for
$F_{D} \ge 0.165$ the stripe state could break apart, resulting in 
a shift
of the peak in $d\langle V_{y}\rangle /dF^{y}_{D}$ to higher $F^y_D$.
Figure~2(b) shows that there is a linear increase in 
$d\langle V_{y}\rangle/dF^{y}_{D}$ 
for $0.1 < F^y_{D} < 0.16$, below the large peak. 
Within this range of $F^y_D$, the steady state flow is strongly fluctuating
as shown in Fig.~3(c) and Fig.~3(d). 
For $0.04 <  F^y_{D} < 0.1$,  
the $\langle V_{y}\rangle$ versus $F_{D}$ 
curve increases very slowly above the depinning transition,
as also indicated by the small value of $dF^{y}_{D}/dF_{D}$
in Fig.~2(b).
In this range of $F^y_{D}$, 
the flow is still plastic and a portion of the particles remain immobile
while others flow past; 
however, the character of the plastic flow differs from the
strongly fluctuating flow found for $0.1 < F^y_{D} < 0.16$. 
The low drive plastic flow takes the form of filamentary flow 
along the stripes, as illustrated in Fig.~3(a)
for $F^y_{D} = 0.05$ where one river of particles
flows along the stripe. 
Figure~3(b) shows that the particles still retain much of the 
stripe structure, in contrast to the strongly fluctuating flow illustrated
in Fig.~3(d) where the stripe structure is nearly lost. 
Another difference is that
the plastic flow in Fig.~3(c) 
involves a significant transverse diffusion of particles 
in the $x$-direction, so that
over time the particles can mix throughout the system.
For the low drive plastic flow in Fig.~3(a), there are
some early time particle jumps transverse to the drive from one stripe 
to another; however, these events vanish in the long
time limit and there is no steady state diffusion in 
the $x$-direction, even though the shape of the filamentary flow within
one stripe may change slightly over time.  

In Fig.~2(c) we plot $\langle V_{x}\rangle$ versus $F^{x}_{D}$ 
for the same system in Fig.~2(a), while the 
corresponding $d\langle V_{x}\rangle /dF^{x}_{D}$ appears in Fig.~2(d).
For the $x$-direction pulse drive,
we do not find any regime where the stripes can dynamically reorient
and align themselves in the $x$ direction.
Instead, the system passes directly into a
moving stripe phase in which the stripe orientation remains perpendicular
to the direction of stripe motion.
The transition into the sliding stripe phase
occurs at the peak in $d\langle V_{x}\rangle/dF^{x}_{D}$ shown in Fig.2(d) at 
$F^x_{D} = 0.224$.  This peak falls at a higher value of $F^x_D$ than
the value of $F^y_D$ of the peak in Fig.~2(b).
This is because stripes moving perpendicular to their orientation are much more
susceptible to breaking apart than stripes moving parallel to their orientation.
A higher pulse driving force reduces the 
effectiveness of the pinning and reduces the tendency for plastic flow,
so a stripe moving perpendicular to its orientation
is stabilized at a higher drive than a stripe moving parallel to its orientation
for the same pinning strength.    

Figure~2(d) also shows a 
smaller maximum in $d\langle V_x\rangle/dF^x_D$ 
near $F^x_{D} = 0.135$, corresponding to a change
in the character of the plastic flow. 
For $0.135 \leq  F^{x}_{D} < 0.224$, the steady state flow is dominated by
strong plastic rearrangements where the stripe structure
completely breaks apart.
In contrast, for $0.06 < F^{x}_{D} < 0.135$, the
flow is more filamentary and is composed of a small 
number of slowly changing channels.  
For $ 0.17 < F^{x}_{D} < 0.22$, we also find a 
metastable effect similar to that shown in Fig.~4 for driving in the 
$y$-direction. 
An ordered stripe can slide perpendicular to the direction of the stripe 
orientation for a period of time which increases with increasing $F^x_D$
before the stripe breaks apart.

\begin{figure}
\includegraphics[width=3.5in]{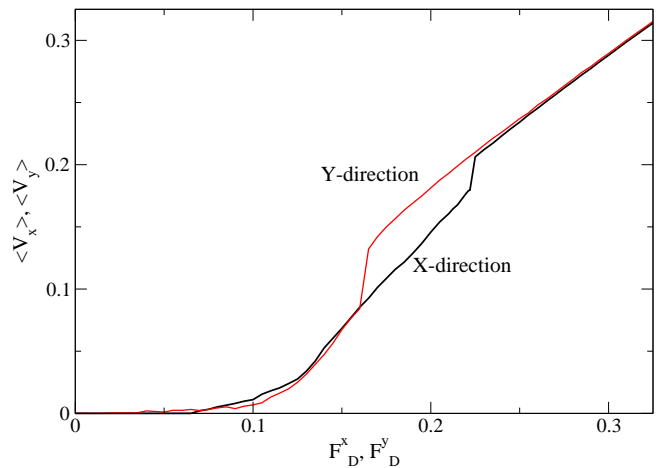}
\caption{
A combined plot of $\langle V_x\rangle$ vs $F_D^x$ (dark line) and
$\langle V_y\rangle$ vs $F_D^y$ (light line)
from Fig.~2(a) and Fig.~2(c) highlights the difference in the 
depinning thresholds $F^x_{C}$ and $F^y_C$ and the transport characteristics.  
}
\end{figure}

In Fig.~5 we plot $\langle V_x\rangle$ versus $F_D^x$ on the same panel
as $\langle V_y\rangle$ versus $F_D^y$
to highlight the regions of anisotropic transport. 
For $0.15 < F^{x,y}_{D} < 0.224$, the velocity is higher for driving along
the easy $y$ direction.
In contrast, for $0.08 < F^{x,y}_{D} < 0.15$, 
$\langle V_{y}\rangle$ falls below $\langle V_{x}\rangle$. 
This occurs because the pulse drive measurements preserve
some of the initial structure of the oriented stripes. 
In the low drive regime $0.08<F^{x,y}_D<0.15$,
much of the flow is filamentary. For driving along the stripe, $F_D^y$,
the filamentary flow settles quickly into a few nonfluctuating channels, 
while for driving against the stripe, $F^x_D$, 
the filamentary flow forms fluctuating plastic channels 
which generally have large velocity pulses, producing a larger average
velocity $\langle V_x\rangle$.          
For $ 0.025 < F^{x,y}_{D} < 0.04$, anisotropy appears due to the differing
critical depinning forces in the two directions; here, 
flow only occurs in the $y$-direction but is absent in the $x$-direction.  

\subsection{Transient Times}

We next analyze in detail the time required for the system to 
achieve steady state flow under a pulse drive.  This time grows rapidly 
near the transition between the disordered and ordered flow states,
as shown in Fig.~4. 
Recent experiments and simulations of periodically sheared particle
assemblies under suddenly applied shear
have shown evidence for a diverging time to reach a steady state
upon approaching a dynamic phase transition \cite{Corte}.
Recent simulations of the plastic depinning of
repulsively interacting particles 
also revealed that the transient time
to reach a steady state under a pulsed drive diverges 
as a power law as the depinning transition is approached from either side
\cite{R}.
This suggests that analyzing the transient time 
required to reach a steady state can be a useful diagnostic for probing
changes in dynamical states.   

\begin{figure}
\includegraphics[width=3.5in]{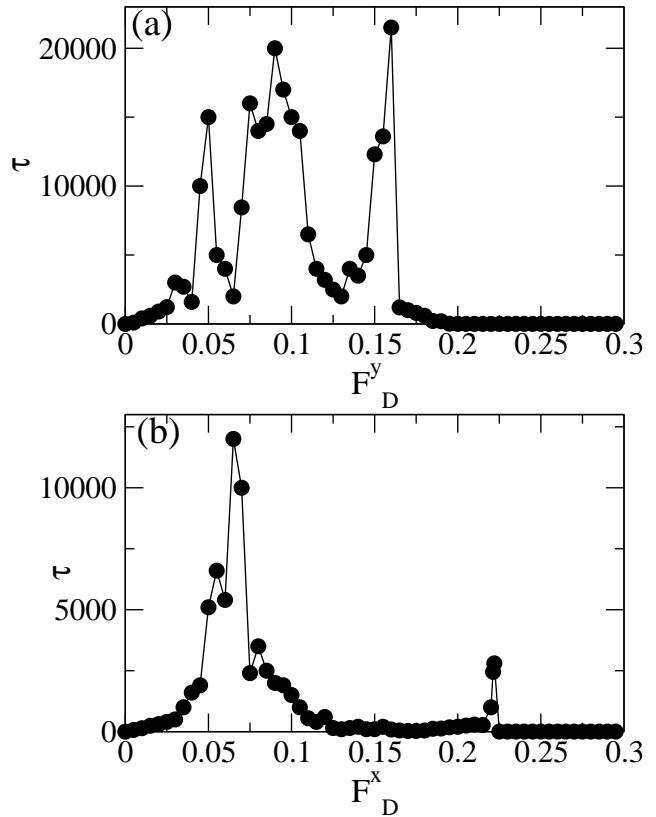}
\caption{
The time $\tau$ required for the system to reach a steady state 
velocity after an applied pulse drive for the system in Fig.~2. 
(a) $\tau$ vs $F^y_D$ for driving in the $y$-direction. 
The first peak at $F^y_D=0.04$ corresponds to the 
depinning transition, the second peak centered at 
$F^y_D=0.1$ appears at the transition from the filamentary flow
illustrated in Fig.~3(a,b) to the fluctuating flow 
state shown in Fig.~3(c,d), and the 
third peak at $F^y_D=0.165$ is associated with the 
transition to a moving ordered stripe state.
(b) $\tau$ vs $F^x_D$ for driving in the $x$-direction.  
}
\end{figure}

In Fig.~6(a) we plot the transient 
time $\tau$ taken by the system in Fig.~2(a) to reach a steady state
after the application of a pulse drive in the $y$-direction. 
For $F^{y}_{D}<0.04$, below the critical depinning force $F^y_c$, there  
is still a finite transient time during which the system organizes into
a pinned state.
The time required to reach the pinned state increases as 
$F^{y}_{c}$ is approached from below, while the time required to reach
a steady moving state increases as $F^y_c$ is approached from above.  
For $F^{y}_{D} \ge 0.165$, when the system passes directly into 
a moving ordered stripe state, the steady state velocity is reached very 
quickly and $\tau$ is very small. 
In the region $ 0.1 < F^{y}_{D} < 0.165$, 
the system depins into a metastable moving ordered stripe state 
that breaks apart after a time $\tau$ as shown in Fig.~4. 
As $F^{y}_{D}$  approaches 
$F^y_D=0.165$ from below, $\tau$ increases 
since it takes increasingly longer times to trigger
the instability that results in the fragmentation of the stripes.
The rapid increase of $\tau$ suggests that $\tau$ may diverge at the 
transition to the moving stripe phase; however, 
our results are not accurate enough to establish
whether this divergence has a power law form. 
We find that the peak in $\tau$ at $F^{y}_{D} = 0.165$ is asymmetric. 
In comparison, the dynamic phase transitions studied in the shearing
systems produced symmetric diverging time scales on both sides of the 
transition \cite{Corte,R}.   
Fig.~6(a) also shows a peak in $\tau$ centered near 
$F^y_D=0.09$, which corresponds to the location of the 
change in slope of $\langle V_y\rangle$ versus $F^{y}_{D}$ in Fig.~2(a). 
At this drive, there is a change from the filamentary plastic flow
channels shown in Fig.~3(a,b) to the rapidly fluctuating disordered 
plastic flow channels shown in Fig.~3(c,d).  
The fact that $\tau$ also increases in this region 
is further evidence that there can be dynamical phase changes
even within the plastic flow regime. 
Finally, there is another peak in $\tau$ near $F^{y}_{D} = 0.04$ at
the depinning transition. 
These results show that peaks in the transient time 
can be used to detect
changes in the flow characteristics of these systems.

In Fig.~6(b) we plot $\tau$ for pulse driving in the $x$-direction. 
Near $F^{x}_{D}=0.225$  
there is a peak in $\tau$ 
associated with the transition 
to the moving perpendicular stripe phase.
Within the range $0.21 < F^{x}_{D} \leq 0.225$, 
where the value of $\tau$ is locally enhanced,
the system forms a metastable state of stripes moving perpendicular
to their length.  Eventually, the stripes
break apart, and the time required for this to occur
increases with increasing $F^{x}_{D}$ 
until for high enough $F^{x}_{D}$ the moving 
perpendicular stripe structure becomes stable rather than metastable
and $\tau$ drops back to a small value.
The peak in $\tau$ at $F_D^x=0.225$ in Fig.~6(b) is 
similar to the peak found for driving in the $y$-direction in Fig.~6(a) near 
$F^{y}_{D} = 0.165$. 
Figure 6(b) shows that
there is another peak in $\tau$ centered at 
$F^{x}_{D} = 0.06$ corresponding
to the transition from filamentary plastic flow to strongly disordered
plastic flow. Overall, the transient times for driving in the 
$x$-direction are smaller than for driving in the $y$-direction. 
For driving along the $y$-direction, 
plastic flow channels can form which do not distort the
aligned stripe pattern, permitting the system to remain in a metastable
state for longer times before falling into the disordered steady state
stripe structure.
For driving along the $x$-direction,
the stripe structure is more strongly disordered 
even at lower drives, so the system is closer to the
disordered steady state stripe structure from the beginning and spends
a shorter amount of time in the metastable state.

\begin{figure}
\includegraphics[width=3.5in]{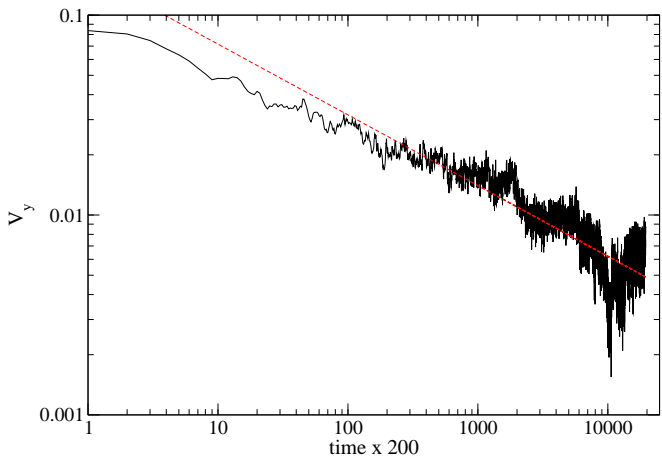}
\caption{
Solid line: $V_{y}$ vs time for $F^{y}_{D} = 0.085$ for the system 
from Fig.~5, showing the decay of the system into the
filamentary plastic flow state.  Dashed line: a power
law fit to $V_{y}(t) \propto t^{-\alpha}$ with $\alpha = 0.35$.  
}
\end{figure}

The filamentary plastic flow regime that
appears for $0.04 < F^y_D < 0.1$ is associated with large values of $\tau$
as shown in Fig.~6(a).
In this regime,
the time decay of $V_y$ to its steady state value differs from the 
decay in the strongly fluctuating plastic flow regime
shown in Fig.~4, where $V_y$ remained roughly constant before dropping
relatively rapidly to a lower value.
Instead, in the filamentary regime $V_y$ follows a continuous
stretched exponential form or power law, 
as illustrated in Fig.~7 for $F_{D}^{y}=0.085$. 
The dashed line is a power law fit to
$V_{y} \propto t^{-\alpha}$ with $\alpha = 0.35$. 
We find an equally good fit of the curve to a
stretched exponential form.
The exponents from the fits do not appear to be universal and change when
we take measurements from the other plastic flow regimes.
Our results indicate that within the 
filamentary plastic flow phases, very long
transient times can occur.

The order-disorder transition at $F^y_D=0.165$ between the 
lower drive strongly fluctuating plastic flow phase shown in Fig.~3(d) 
and the higher drive
moving ordered stripe phase shown in Fig.~3(f)
exhibits metastability and has a diverging time scale only on
the low drive side of the transition, as indicated in Fig.~6(a).
These features strongly suggest that this transition is first order in nature
and that the details of the transition are strongly affected by the initial
conditions of the moving state.  For example, it is possible to obtain
a reversed metastability by
starting
the system in a disordered configuration and applying a pulse drive
$F^y_D>0.165$.
In this case, 
the moving system remains disordered and travels at a lower velocity 
until an instability causes the stripe structure to form with a corresponding
increase in the velocity.
This reversed metastability shows diverging transient times as the
transition is approached from above, but has no diverging time scales
when the transition is approached from below.
The metastability of the ordered and disordered states 
resembles the superheating or supercooling recently observed for
systems with first order phase transitions. 
Very similar dynamical superheating and supercooling effects were found
in superconducting vortex systems in
experiments \cite{Andrei} and three-dimensional (3D) 
simulations \cite{Scalettar}.
The vortex system undergoes a disorder-induced first order phase 
transition, and the effective disorder changes
when the system is prepared in different states. 
Computational studies of 2D vortex systems 
interacting with disorder have shown
that there is either a continuous order to disorder transition or 
a crossover, so hysteresis, superheating and supercooling do not appear.
For many stripe forming systems in 
two dimensions, transitions from ordered to disordered
states in equilibrium and in the absence of quenched disorder 
are first order in nature \cite{Pell,Glaser}. 
Our results suggest that the first order nature of the 
equilibrium transitions persists 
for some of the transitions between nonequilibrium states.  

At the transition between the low drive filamentary plastic flow phase and
the higher drive strongly fluctuating plastic flow phase,
Fig.~6(a) indicates that there are diverging transient times on both
sides of the transition.  This behavior is similar to the diverging
transient times found for
2D plastic depinning.  There is evidence that the plastic depinning
is an absorbing phase transition falling in the 
directed percolation class \cite{R}. 
Determining whether the depinning of the stripe system or the
filamentary plastic flow to strongly fluctuating plastic flow transition
are also nonequilibrium phase transitions falling in the directed percolation
class is beyond the scope of this work;
however, our results suggest that the stripe system 
may be an ideal system in which to examine the nature of nonequilibrium
transitions since it exhibits several different types of flow phases. 

\begin{figure}
\includegraphics[width=3.5in]{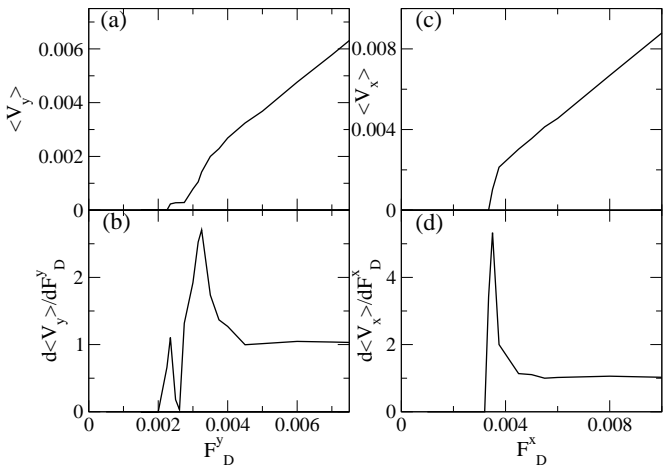}
\caption{
(a) $\langle V_{y}\rangle$ vs $F^{y}_{D}$ 
for a pulse drive system with $\rho_p=0.38$ and a lower pinning force of
$F_{p} = 0.125$. 
A two step depinning occurs, with uncoupled stripes depinning initially
followed by the depinning of coupled stripes.
(b) The corresponding
$d\langle V_{y}\rangle/dF^{y}_{D}$ 
has a double peak indicating the two step depinning process.
(c) $\langle V_{x}\rangle$ vs $F^{x}_{D}$ 
for the same system. 
(d) The corresponding $d\langle V_{x}\rangle/dF^{x}_{D}$ shows
a single step elastic depinning of the stripes.  
}
\end{figure}

\begin{figure}
\includegraphics[width=3.5in]{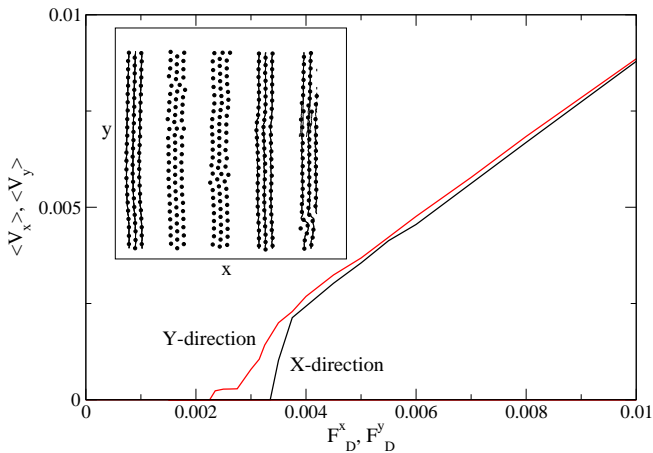}
\caption{
A combined plot of $\langle V_x\rangle$ vs $F_D^x$ (dark line) and
$\langle V_{y}\rangle$ vs $F_{D}^y$ (light line) 
for the system in Fig.~8 
with $F_p=0.125$ highlighting the transport anisotropy.  
Inset: The particle positions (dots) and 
trajectories (lines) for the system in Fig.~8(a) in the
decoupled stripe regime at $F^y_D=0.0025$
where a portion of the stripes are moving while others are pinned. 
}
\end{figure}

\subsection{Pulse Measurements for Weak Disorder}

We next consider pulse drive measurements for a system with
the same pinning density $\rho_p=0.38$ but with a weaker disorder
strength of $F_p=0.125$.
In Fig.~8(c) we plot $\langle V_{x}\rangle$ versus $F^{x}_{D}$ 
and in Fig.~8(d) we show the corresponding $d\langle V_{x}\rangle/dF^{x}_{D}$. 
For this value of $F_{p}$, the stripe
structure remains ordered. 
The single step depinning for $x$-direction driving is elastic 
and has $\langle V_x\rangle \propto (F^x_{D} - F^x_{c})^\beta$, 
with $\beta =  0.35$.  
This is followed by a crossover to $\langle V_{x}\rangle \propto F^{x}_{D}$
at higher drives. 
The behavior agrees well with
the depinning of a 
harmonic elastic string driven over a random substrate, where 
an exponent $\beta = 0.33$ is observed
\cite{Krauth}.
For driving in the $y$-direction, as shown in Fig.~8(a), 
$\langle V_{y}\rangle$ versus $F_{D}$ indicates that
a two step depinning process occurs.
The initial depinning involves the flow of individual stripes, while other
stripes remain pinned or move at different velocities, as illustrated
in the inset of Fig.~9 for $F^y_D=0.0025$.
At higher drives, the remaining stripes depin and become coupled 
to the other moving stripes, resulting in an
elastic flow. 
The two peaks in $d\langle V_y\rangle/dF^y_D$ shown in Fig.~8(b) fall at
the locations of the two depinning transitions.
A similar type of two step, layered depinning transition 
was predicted for anisotropic charge 
density wave (CDW) systems, 
where CDWs first depin separately and flow independently from one layer
to the next, and then recouple at higher drives \cite{Vinokur}.
Mean field models also predict that layered systems
should show a coupling-decoupling transition \cite{M2,Saunders}, 
while two-layer models
predict that coexistence of moving and pinned phases should occur in 2D
systems
\cite{MC}. 
In the main panel of Fig.~9, we plot $\langle V_x\rangle$ versus
$F^x_D$ and $\langle V_y\rangle$ versus $F^y_D$ together
in order to highlight the
transport anisotropy 
which disappears for $F^{x,y}_{D} > 0.04$ when fully elastic flow
is established.    
The transient times 
$\tau$ for the $F_p=0.125$ system are much shorter than those
in the $F_p=0.9$ system, where plastic depinning occurred.
We find an increase in $\tau$ just below each depinning transition in the
$F_p=0.125$ system for both $x$ and $y$ direction driving.
For $x$-direction driving, 
there is a single peak in $\tau$ below the depinning threshold 
which is associated with a small amount of roughening of the stripe structure
that occurs just before depinning.
Above each depinning transition
in the weak pinning system, $\tau$ is extremely small.
For $y$-direction driving, 
the peak in $\tau$ is broader within the sliding plastic flow phase found
below the second depinning transition.
In systems with even weaker pinning, $F_{p} < 0.05$, 
the depinning is elastic for both driving directions 
and the transport anisotropy is significantly reduced.

\begin{figure}
\includegraphics[width=3.5in]{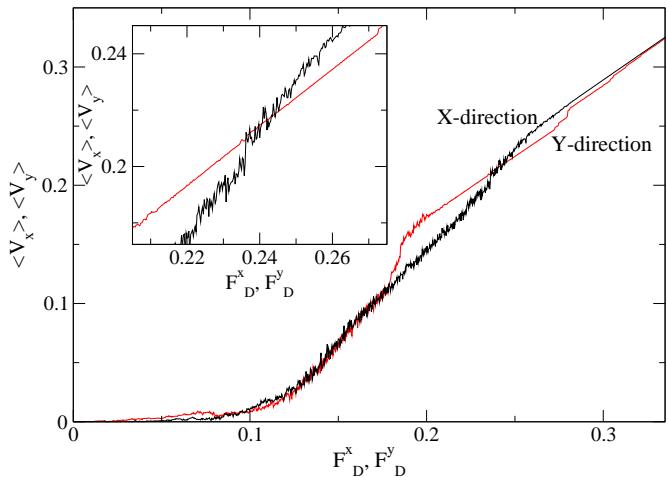}
\caption{
A combined plot of $\langle V_{x}\rangle$ 
vs $F^x_D$ (dark line) and $\langle V_{y}\rangle$ vs $F^y_D$ (lighter line)
for a system with $F_p=0.9$ and the same parameters as in Fig.~2 
but with a continuously swept drive.
There are several steps in $\langle V_{y}\rangle$ 
that appear below the drive $F^{x,y}_D=0.32$ at which
the two curves meet.
Inset:  A blow-up of the region near $F^{x,y}_{D} = 0.24$ shows
a crossing of the $\langle V_x\rangle$ and $\langle V_y\rangle$ curves, 
indicating that $\langle V_y\rangle$ has a smaller slope and that
therefore fewer particles are moving for
driving in the $y$-direction than for driving in the $x$-direction. 
The crossing  of $\langle V_x\rangle$ and $\langle V_y\rangle$
does not occur for the pulse drive measurements shown in Fig.~5. 
}
\end{figure}

\begin{figure}
\includegraphics[width=3.5in]{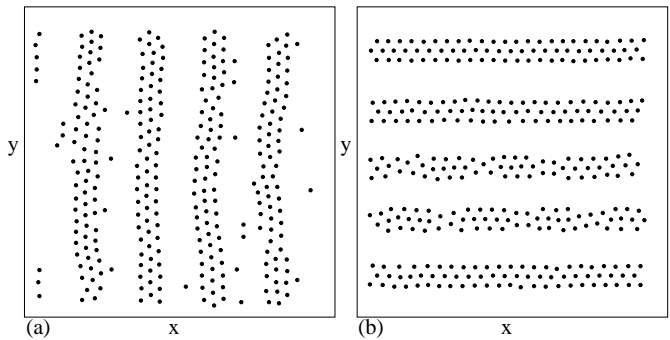}
\caption{
(a) Image of particle positions in the reordered phase at $F^y_D=0.27$ in the
swept drive system from Fig.~10.  In addition
to the stripe structure, there are some scattered particles pinned 
between the stripes. 
(b)  Image of particle positions in the reordered phase at $F^x_D=0.27$ in the
same system.  The stripes have reoriented in the $x$-direction
and there are no pinned particles between the stripes.
}
\end{figure}

\section{Continuous Force Sweep Measurements}

\subsection{Strong Disorder}

We now examine the case where the applied drive 
is slowly incrementally increased in a single sweep as opposed 
to the sudden application of the drive discussed in the previous section. 
We use a driving force increment of $\Delta F_D=0.00025$ applied
every 25000 simulation time steps in a system with
$F_{p} = 0.9$ and 
$\rho_{p} = 0.38$. 
In Fig.~10 
we plot the resulting $\langle V_{x}\rangle$ versus $F^x_D$ and 
$\langle V_{y}\rangle$ versus $F^y_D$ together.
Many of the transport features are the same as 
those shown for the pulse drive in Fig.~5, 
such as the lower depinning threshold
for driving in the $y$-direction and the 
anisotropic flow centered near $F^{x,y}_{D} = 0.2$.
The transition to the flowing stripe state remains sharp for the continuous
sweep drive in the $y$-direction, but for the $x$-direction driving
the state in which the stripes flow perpendicular to their orientation is
lost.
Instead, the system reorders into a stripe state oriented
along the $x$-direction, as shown in Fig.~11(b) for $F^{x}_{D} = 0.27$. 
In the pulse drive system, the perpendicular moving 
stripe state can be stabilized by dynamical quenching, 
but for the swept drive, the system passes through an 
extensive plastic flow phase which destroys 
any memory of the initial perpendicular stripe orientation. 

The inset of Fig.~10 highlights that under the swept drive,
a crossing of $\langle V_{y}\rangle$ and  $\langle V_{x}\rangle$ 
occurs near $F^{x,y}_{D} = 0.24$.  There is no such crossing for
the pulse drive, as shown in Fig.~5. 
The crossing of the curves indicates that although the system is in
the moving stripe state for both drive directions, the slope of 
$\langle V_y\rangle$ is smaller than that of $\langle V_x\rangle$ in this
regime.  This is because a fraction of the particles remain pinned for
the $y$-direction drive, while all of the particles are moving for the
$x$-direction drive.  When fewer particles are moving, the slope of the
velocity-force curve is reduced.  Particles are able to remain pinned for
the $y$-direction drive because they can be captured in pinning sites that
are sufficiently far away from the neighboring stripes that they experience
only repulsion from the particles in the stripes, and are out of 
the range of the
attractive part of the particle-particle interaction potential.  This does
not happen for the $x$-direction drive because during the stripe reorientation
process, all pinned particles are eventually swept up into a moving stripe.
In Fig.~11(a), the reordered stripe phase for $F^{y}_{D} = 0.27$ 
contains pinned particles that sit between the stripes rather than flowing
with the stripes.
In contrast, in the moving stripe phase for a $y$-direction pulse drive shown
in Fig.~3(f), there are no pinned particles between the moving stripes. 
In the $x$-direction swept drive moving stripe phase, all of the particles
are moving and there are
no pinned particles between the moving stripes, as shown in
Fig.~11(b) for $F^x_D=0.27$.
The transition into the moving stripe phase
for swept $y$-direction driving is rapid, as indicated by the jump
in $\langle V_y\rangle$ at $F^y_{D} = 0.185$ 
in Fig.~10.  
For $x$-direction driving, the
reordering transition is more continuous, permitting more meandering
of the stripe pattern during the stripe formation process. 
This allows all of the pinned particles to be attracted 
gradually into the
moving stripe structure. 
     
Once the driving force becomes strong enough,
the pinned particles surrounding the moving stripes for the $y$-direction
swept drive depin and join the moving stripe structures.
The depinning of the individual particles produces
the step features in $\langle V_{y}\rangle$ near $F^y_{D} = 0.28$ 
in Fig.~10. 
For drives above the depinning threshold of all of the pinned particles,
the velocity response is isotropic, as shown in Fig.~10 for 
$F_D^{x,y}>0.32$.
Our results indicate that even for pinning strengths strong enough to induce
plastic flow and a subsequent reordering of the stripe structure, 
some memory of the initial ordering of the stripe phase is
retained up to relatively large values of the driving force.
We note that based on the extended transient behavior found in the plastic
flow regimes for the pulse drives, it is possible that if the swept drives
were applied with even slower drive increments, the anisotropic response
could be lost at lower drives if the system is given more time to slowly 
mix in the plastic flow state.

\begin{figure}
\includegraphics[width=3.5in]{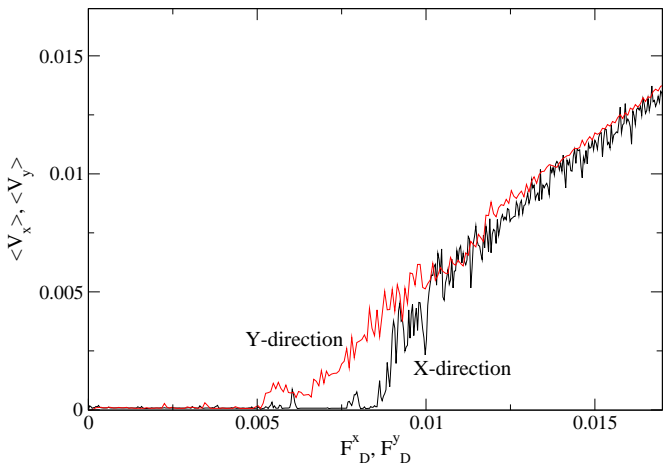}
\caption{
A combined plot of $\langle V_x\rangle$ vs $F^x_D$ (dark line) and
$\langle V_y\rangle$ vs $F^y_D$ (lighter line) for a system with
$\rho_p=0.38$, a lower $F_p=0.225$, and a swept drive.
The depinning in the $x$-direction is accompanied
by plastic distortions and the partial breaking apart of the stripe
structure. 
Depinning in the $y$-direction occurs 
by the sliding of stripes past one another 
in a manner similar to that illustrated
in the inset of Fig.~9.
}
\end{figure}

\begin{figure}
\includegraphics[width=3.5in]{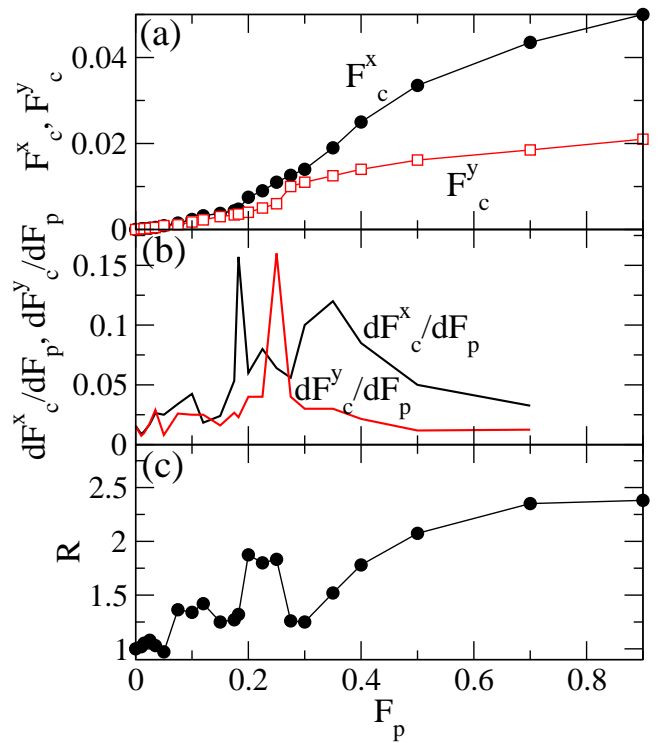}
\caption{
(a) The critical depinning forces $F^x_{c}$ (filled circles) and 
$F^y_c$ (open squares) for driving in the $x$ and $y$-directions,
respectively, plotted vs $F_{p}$.
The sharp jumps in the depinning forces are associated with
changes from elastic flow to plastic flow and
from plastic flow to sliding ordered flow.
(b) The corresponding $dF^{x}_{c}/dF_{P}$ vs $F_p$ (dark line) and
$dF^{y}_c/dF_P$ vs $F_p$ (light line)
curves more clearly show the onset of the different
depinning phases. 
(c) The anisotropy ratio 
$R = F^{x}_{c}/F^{y}_{c}$ vs $F_p$ shows that the change in anisotropy
can be associated with different depinning regimes.
}
\end{figure}

\subsection{Pinning Strength Dependence and Peak Effect}

When we perform swept drive measurements on a system with the weaker pinning
strength of $F_p=0.125$, we obtain velocity-force curves that are nearly
identical to those shown in Fig.~8 for a pulse drive measurement.
We attribute this to the lack of plastic flow in the weakly pinned system.
Without plastic flow, the stripe structure never breaks apart and memory of
the initial stripe orientation is never lost, so the same stripe orientation
appears, regardless of whether the drive is slowly swept or suddenly applied.
By increasing $F_p$ slightly, we reach a state where the $x$-direction depinning
is plastic and accompanied by the breaking apart of the stripe structure, while
the $y$-direction depinning occurs by the sliding of some of the stripes, with
the stripe structure maintained intact.  This situation is illustrated in
Fig.~12, where we plot $\langle V_x\rangle$ versus $F^x_D$ and 
$\langle V_y\rangle$ versus $F^y_D$ for a swept drive system with
$F_p=0.225$.

By conducting a series of simulations for 
varied $F_{p}$, we map the anisotropy in the depinning thresholds
$F^x_c$ and $F^y_c$, as shown in Fig.~13(a).
For $x$-direction driving, 
$F^{x}_{c}$ increases monotonically with increasing $F_p$ 
for $ 0 < F_{p}< 0.175$. 
Within this range of $F_p$, the depinning is elastic and the stripes move
perpendicularly to their orientation.  Each particle maintains its position
in its original stripe.
Just above $F_{p} = 0.175$, there is a sudden
increase in $F^{x}_{c}$ corresponding to the onset 
of plastic distortions of the stripe structure at depinning. 
For driving in the $y$-direction, $F^{y}_{c}$ 
continuously increases with increasing $F_p$ for 
$0 < F_{p} < 0.25$.  Over this range of $F_p$,
the stripes depin elastically and either all depin simultaneously
for $ F_{p} < 0.05$, or depin as individual sliding stripes for
$0.05 \leq F_p < 0.25$ via the mechanism illustrated in the inset of
Fig.~9.
For $F_p<0.05$, where our system behaves elastically, we find
that $F^{y}_{c}$ increases approximately quadratically
with $F_{p}$ 
which is the expected behavior for elastic depinning \cite{Brandt}.   
Just above $F_{p} = 0.25$ there is a sharp increase in $F^{c}_{y}$ when the 
stripes begin to depin plastically along individual 
stripes in the manner shown in Fig.~3(a). 

For $F_p>0.4$, both $F^{y}_{c}$ and $F^{x}_{c}$ begin to saturate, as
shown by the anisotropy ratio $R=F^{c}_{x}/F^{c}_{y}$ 
plotted in Fig.~13(c). 
The saturation arises due to the fact that the number of pinning sites $N_p$ is
only slightly higher than the number of particles $N$.  As a result, when
$F_p$ is large enough, the initial depinning is dominated by interstitially
pinned particles.  These particles are not trapped by one of the randomly
located pins, but are instead held in place by interactions with
neighboring pinned particles.  The critical force for the depinning of 
particles trapped by pins increases linearly with increasing $F_p$, but the
critical force for the depinning of interstitially pinned particles is
determined only by the particle-particle interaction potential 
and is not altered by increasing $F_p$.
In a system with a much higher pinning density, every particle would be
trapped by a pin and the depinning threshold would show the expected linear
increase with increasing $F_p$.

The sharp increases in the depinning thresholds 
associated with transitions from elastic to plastic flow or from 
weakly plastic flow to a more strongly plastic flow 
resemble the phenomenon observed for depinning
of vortices in type-II superconductors, 
where a peak in the depinning threshold has been connected with the
disordering of the vortex lattice 
\cite{Higgins,H,Peak,Ling,P,Moh,Andrei,Scalettar}. 
In the disordered or plastically flowing systems, the particles
can more readily adjust their positions to take advantage of the 
energy of a randomly located pinning site without paying the large 
energy cost required to distort an elastic or ordered particle lattice.
In studies of 2D vortex systems, as the vortex lattice
is softened the system becomes more disordered 
and the depinning threshold increases continuously, 
producing a peak effect that is continuous rather than sharp 
\cite{SB}.  This may be 
related to the fact that 2D systems of particles with repulsive 
long range interactions lack first order melting or disordering transitions.
Many 3D vortex systems have a first order transition from an ordered 
vortex structure to a disordered one, 
and the peak effect phenomena observed in these
systems is very sharp. 
Even through the stripe system described here is 2D, the pairwise 
particle interactions
are not strictly repulsive. 
Previous studies in the absence of quenched
disorder using the same model produced results 
that suggest that the thermal melting of the stripe and clump 
systems is a first order transition. 
Other 2D studies of stripe-forming systems with competing
interactions also find first order melting 
transitions for many of the phases \cite{Pell}.
These results suggest that the first order 
peak effect phenomenon found for vortex systems 
may also generically occur for stripe
and pattern forming systems in the presence of quenched disorder     

To further characterize the changes in $F_{c}$ we 
plot $dF^{x}_{c}/dF_{p}$ and $dF^y_c/dF_p$ vs $F_p$ in Fig.~13(b). 
There is a peak in $dF^x_c/dF_p$
near $F_{p} = 0.185$ corresponding to the transition from ordered
stripe flow to partially plastic flow.
A second, much broader peak appears
near $F_{p} = 0.35$ at the point where the stripes break apart completely.
For $0.185 < F_{p} < 0.35$, the stripes driven in the $x$-direction 
reorder into a perpendicularly moving stripe state at high $F^x_D$ in
spite of the fact that weakly plastic flow occurs above depinning.
For $F_{p} \geq 0.35$, the plasticity at depinning becomes much stronger and
the stripes reorder into a parallel moving state at high $F^x_D$. 
For driving in the $y$-direction we find a single sharp peak 
in $dF^y_c/dF_p$
near $F_{p}=  0.25$ corresponding to the appearance of
plastic flow along the stripes. 
We note that the transition 
at $F_p=0.075$ into the pinned-sliding phase
illustrated in the inset of Fig.~9 
is not associated with any sharp features in $dF^{y}_{c}/dF_{p}$.       

The appearance of different flow phases can also be detected in the plot of
the anisotropy ratio $R$ in Fig.~13(c). 
For example, the peak in $R$ over the range 
$0.185 < F_{p} < 0.3$ occurs when the flow for $x$-direction driving is
plastic, while the flow for $y$-direction driving remains elastic in the
stripe sliding state.
The increase in $R$ at $F_{p} = 0.06$ 
corresponds to the transition from elastic depinning in both directions to
elastic depinning for $x$-direction driving and individual stripe sliding
for $y$-direction driving.
For $F_{p}>0.6$, the anisotropy begins to saturate.

\begin{figure}
\includegraphics[width=3.5in]{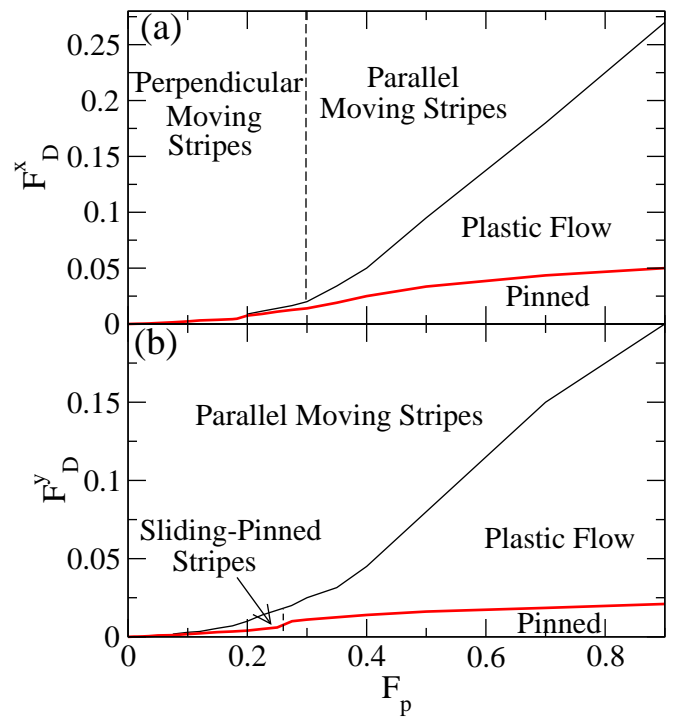}
\caption{
Dynamic phase diagrams 
showing the onset of depinning (lower heavy line) 
and the onset of the dynamic stripe reordering
(upper light line). 
(a) $F^{x}_{D}$ vs $F_{p}$. 
The dashed line separates the moving perpendicular stripe phase from the 
dynamically reordered parallel stripe phase. 
In order for the parallel stripe phase to form, the system must first
pass through a strongly fluctuating plastic flow phase.
(b) $F^y_D$ vs $F_p$.
The dashed line indicates the change from sliding stripe plastic flow 
to plastic flow in which the stripe structure breaks apart. 
}
\end{figure}

\subsection{Dynamic Phase Diagram}

By identifying features in the velocity force curves 
and associating them with different moving states, 
we construct a dynamic phase diagram
for both $x$ and $y$ direction driving for $F_{p} = 0.225$. 
In Fig.~14(a) we plot the phase
diagram for $F^{x}_{D}$ vs $F_{p}$, indicating the location of the
depinning curve and the transition into an ordered stripe state.
For  $F_{p} < 0.185$, the stripes depin elastically 
into a perpendicular moving stripe state, while for $0.185 \leq F_p < 0.35$,
the stripes depin with a small amount of plastic distortion into the same
perpendicular moving stripe state. 
The line in Fig.~14(a) marking the transition from plastic flow to  
reoriented ordered parallel moving stripes increases roughly linearly
with $F_p$ for $F_p > 0.4$.

In Fig.~14(b) we plot the phase diagram of $F^y_D$ versus $F_p$ 
for driving in the $y$-direction. 
In this case the stripes are always oriented in the direction of the
drive.
The range of the plastic flow regime grows with increasing 
$F_{p}$ and in general
the onset of the moving stripe phase occurs at lower drives than those
at which the parallel moving stripes form for $x$-direction driving.
The small dashed line indicates the transition from the plastic flow in which
the stripe structure is destroyed
for $F_p \geq 0.275$ to the state where moving stripes slide past pinned
stripes 
for $0.075 < F_{p} < 0.275$. 

The dynamic phase diagram for the stripe 
system contains a larger number of phases 
than dynamic phase diagrams observed in 
systems with purely repulsive particle-particle interactions moving over
random disorder.
For example, in 2D vortex systems the dynamic phases consist only of
a pinned state, a plastic flow state, and a moving partially ordered state,
and the transitions between these states are continuous. 
In the partially ordered moving state, 
the particles are not fully crystallized 
but develop a smectic type of ordering and flow in evenly spaced
channels aligned with the direction of the drive.  The channels of flow
may be coupled or partially coupled
\cite{Koshelev,Balents,Zimanyi,Olson2,Marchetti}. 
This is similar to the state we observe in which the stripes reorient
in the direction of drive for sufficiently high drive and sufficiently strong
pinning.
In the stripe system, 
the stripe reordering transition is more consistent with a 
first order phase transition rather than the continuous or crossover behavior
found for particles with purely repulsive interactions.  
Studies of driven systems with quenched disorder, 
where the particle-particle interactions are more complicated than the
purely repulsive case, have shown that it is possible to have a
coexistence of different moving phases, which is consistent with having
first order phase transitions between the moving phases \cite{Menon}.

\begin{figure}
\includegraphics[width=3.5in]{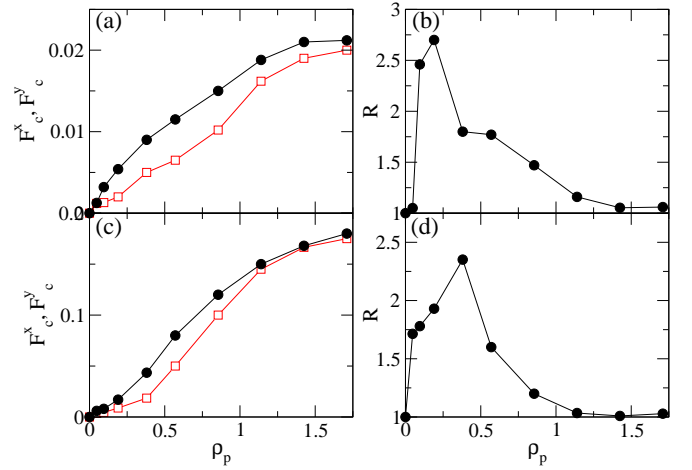}
\caption{
(a) $F^{x}_{c}$ (filled circles) and $F^{y}_{c}$ (open squares) 
vs $\rho_{p}$ for a system with $F_{p} = 0.225$ and $R_{p} = 0.3$. 
(b) The corresponding 
$R = F^{x}_{c}/F^{y}_{c}$ vs $\rho_{p}$ shows
that at large $\rho_{p}$ the anisotropy is reduced.   
(c) $F^{x}_c$ (filled circles) and $F^y_c$ (open squares)
vs $\rho_p$ for a system with $F_{p} = 0.7$ and $R_p=0.3$. 
(d) The corresponding $R$ vs $\rho_p$.  Here the anisotropy drops
nearly to $R=1$ at high $\rho_{p}$    
}
\end{figure}
 
\subsection{Changing Disorder Density and Radius }    

In Fig.~15(a) we plot $F^x_{c}$ and $F^y_c$ versus 
$\rho_{p}$ for a system with $F_{p} = 0.225$ and $R_p=0.3$,
and in Fig.~15(b) we show 
the resulting anisotropy ratio $R = F^{x}_{c}/F^{y}_{c}$ versus $\rho_p$. 
Here $F^x_{c}$ and $F^y_c$ both increase 
monotonically with increasing $\rho_{p}$. 
At low $\rho_{p}$ the
anisotropy $R$ is strongly reduced and the 
two depinning curves come together when the
depinning becomes elastic for both $x$ and $y$-direction drives. 
A similar effect appears in Fig.~13(c) for low $F_{p}$, where the onset
of elastic depinning for both driving directions results in a reduced
value of $R$.
The anisotropy in Fig.~15(b)  
passes through a maximum near $\rho_{p} = 0.2$ before
gradually falling back to $1.0$ for increasing $\rho_{p}$. 
The pinning density plays a more important role in determining the anisotropy
of the depinning for pre-formed stripe states than the pinning strength.
When the average distance between pinning sites 
is greater than the inter-stripe distance, 
depinning for $y$-direction driving in a system with strong pinning
occurs via a combination of plastic flow of particles along 
some stripes while other stripes remain completely pinned.
As a result, $F^y_c$ is generally lower than $F^x_c$ for the low density, 
strong pinning limit.
This is shown in Fig.~15(c) where
we plot $F^{y}_{c}$ and $F^{x}_{c}$ versus $\rho_{p}$ for $F_{p} = 0.7$.
For higher values of $\rho_{p}$, the plastic depinning
along the stripes for $y$-direction driving is suppressed and the
behavior becomes more isotropic. 
For systems with small $\rho_{p}$ and $F_{p} > 0.35$, the
anisotropy persists down to much lower values of 
$\rho_{p}$ than shown in Fig.~15(a); however,
for $\rho_{p} > 1.0$ the anisotropy vanishes completely. 

\begin{figure}
\includegraphics[width=3.5in]{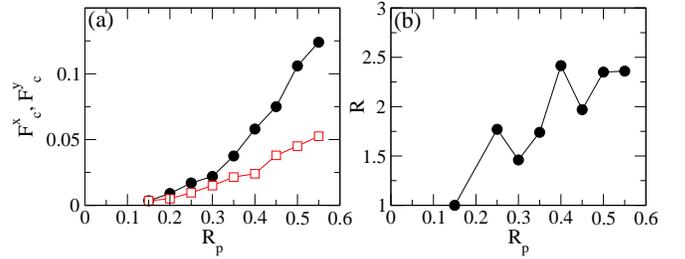}
\caption{
(a) $F^{x}_{c}$ (filled circles) and $F^{y}_{c}$ (open squares) 
vs $R_{p}$ for a system with $F_{p} = 0.225$ and $\rho_{p} = 0.38$.
(b) The ratio $R = F^{x}_{c}/F^{y}_{c}$ vs $R_p$
showing that the anisotropy increases 
with increasing $R_p$ and saturates at high $R_{p}$.   
}
\end{figure}

We next consider a system with $F_{p} = 0.225$ 
and $\rho_{p} = 0.38$ with varied
pinning radius $R_{p}$. 
We plot $F^x_c$ and $F^y_c$ versus $R_p$ in Fig.~16(a), and the anisotropy
ratio
$R = F^{x}_{c}/F^{y}_{c}$ in Fig.~16(b). 
Both $F^x_c$ and $F^y_c$ increase with increasing $R_p$.
For low $R_p$, $F^{x}_{c}$ increases faster with increasing $R_p$ than
$F^y_c$ does, and for high $R_p$, the anisotropy $R$ saturates.

\section{Velocity Fluctuations}

The dynamic phases can also be characterized by measuring the velocity
fluctuations.
Experiments on stripe forming systems 
previously demonstrated that narrow band noise, characterized
by a periodic noise signal, and broad band noise, which lacks 
any characteristic frequencies, 
can occur and showed evidence for transitions between the different
types of noise \cite{N,GT}.
Previous simulations of clump and stripe forming systems 
showed the presence of a $1/f$ noise characteristic in the nonlinear portion
of the velocity-force curve associated with the fluctuating plastic flow
phase.  At high drives where the stripes or clumps reorder, 
the noise becomes white with a weak narrow band or 
washboard frequency similar to that
observed for 
a driven vortex lattice in the dynamically reordered regime \cite{Olson2}.  

\begin{figure}
\includegraphics[width=3.5in]{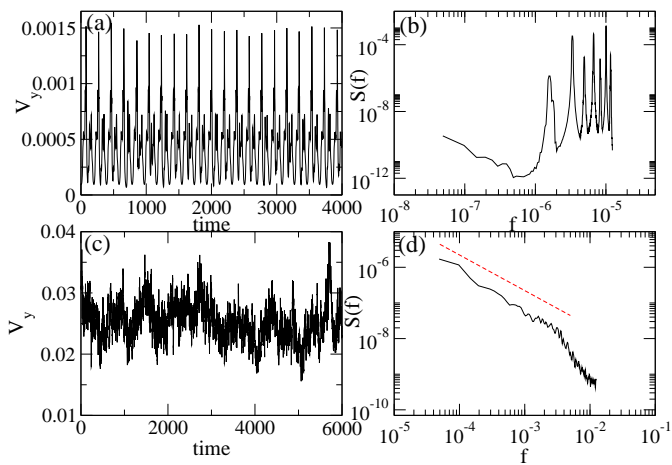}
\caption{
(a) The time series of $V_y$ 
for a system with
$F_{p} = 0.9$, $\rho_{p} = 0.36$, $R_{p} = 0.2$, and 
$F^{y}_{D} = 0.04$. 
For these parameters, the particles move in stationary plastic filaments,
producing a periodic time-of-flight signal.
(b) The power spectrum $S(f)$ of 
the time series in (a) 
showing the characteristic peaks from the periodic signal.
Here, $f$ is in units of inverse simulation time steps.
(c) Time series of $V_y$ for the same system at 
$F^y_{D} = 0.12$ in the strongly fluctuating plastic 
flow regime. (d) The 
corresponding $S(f)$ has a $1/f$ 
feature at low frequencies as indicated by the dashed line.    
}
\end{figure}

Here we show that the stripe system exhibits many 
additional noise features near depinning.  For $y$-direction driving
near depinning,
there can be filamentary plastic flow channels along the stripe with
no particle diffusion from stripe to stripe, as shown in Fig.~3(a). 
Within this filamentary plastic flow regime,
it is possible for the particle flow to be limited to
one or a small number of individual winding channels 
which do not change over time and which have a characteristic time-of-flight
for crossing the sample.
This results in a periodic velocity signal such as 
that shown in Fig.~17(a) for a system
with $F_{p} = 0.9$ and $F^{y}_{d} = 0.04$. 
Similar periodic 
filamentary plastic motion has 
been observed in vortex simulations performed just at depinning. 
Here one or
two stable channels of moving particles form while the rest of the particles
are immobile \cite{Dom}. 
Evidence for filamentary flow has also been found in 
vortex experiments, 
where a series of jumps and dips in the current-voltage curve 
were interpreted as indicating the opening of individual channels of
vortex flow \cite{Am}.
There are also several simulations of vortex systems
showing transitions from narrow band filamentary flow 
to chaotic flow as the drive is increased \cite{Soret}.  
The power spectrum $S(f)$ of the 
velocity time series in Fig.~17(a) is shown in Fig.~17(b), 
and has characteristic narrow band noise peaks produced by the 
time-of-flight signature.
As we increase $F^y_{D}$ and permit the system to settle into a steady state, 
we observe a seres of transitions from ordered flows to 
fluctuating flows with broad band noise signatures.
This behavior very similar to that found in 2DEG transport 
measurements \cite{N}. 
We observe filamentary plastic flow from the depinning transition up to
$F^y_D=0.1$.  Above this drive, the system transitions into the strongly
fluctuating plastic flow regime 
shown in Fig.~3(c) in which the stripe structure is destroyed.
In Fig.~17(c) we plot the time series of $V_y$ for the same system in
Fig.~17(a) at a drive of $F^y_D=0.12$ in the strongly fluctuating plastic
flow regime.  The corresponding $S(f)$ appears in Fig.~17(d).
Here, $V_y(t)$ is strongly fluctuating since the number of pinned particles
in the system is continuously changing,
and the power spectrum shows a $1/f$ noise characteristic 
at low frequencies. 
As the drive is further increased, the system
reorders into a moving stripe state and the
low frequency $1/f$ spectral signal is lost. 
It is replaced by a weak narrow band noise signal similar to that
observed in previous simulations for stripe reordering \cite{Olson2}.     

For driving in the $x$-direction at $F_{p} = 0.9$,  
the filamentary motion along the stripes of the type
seen for driving along the $y$-direction is strongly suppressed.  
Instead, the strongly fluctuating plastic flow states with 
$1/f$ noise signatures are more prevalent. 
When the stripes reorder by reorienting into the $x$-direction, 
we find the same weak narrow band noise feature observed above the
stripe reordering transition for driving in the $y$-direction.

\begin{figure}
\includegraphics[width=3.5in]{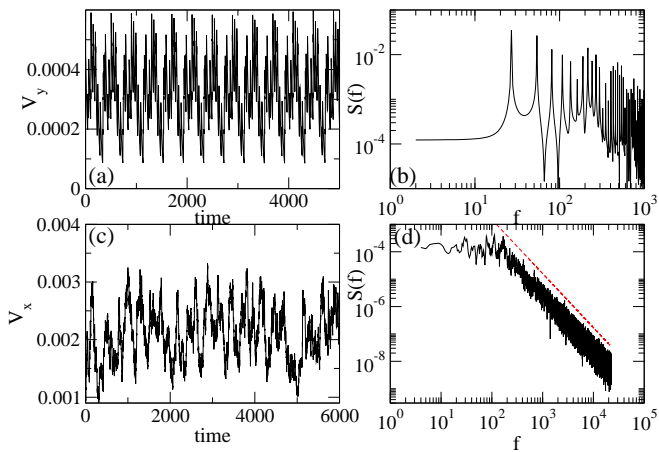}
\caption{
(a) The time series $V_y$ for a system with $F_{p} = 0.125$ and
$F^{y}_{D} = 0.0025$, in the sliding phase. 
(b) The corresponding power spectrum $S(f)$. 
A strong narrow band noise signal appears.
(c) Time series $V_x$ for the same system under 
driving in the $x$-direction at $F^{x}_{D} = 0.0375$, just above depinning.
(d) The corresponding $S(f)$.  
The velocity signal is not periodic; however, 
the power spectrum in (d) has a Lorentzian shape
with a flat spectrum at lower frequencies and a 
$1/f^{2}$ spectra at higher frequencies, indicated by the dashed line. 
For high values
of $F^{x}_{D}$, the power spectrum develops a narrow band noise characteristic. 
}
\end{figure}

Near depinning in a system with weaker pinning of $F_p=0.125$,
the sliding stripe phase produces a periodic signal as shown in 
Fig.~18(a) for $F^{y}_{D} = 0.0025$. 
The periodic signal also appears in the corresponding power spectrum
$S(f)$ shown in Fig.~18(b).
In this regime, the stripes are decoupled, so individual stripes are moving
at slightly different velocities.  This results in a more complex velocity
signal composed of several similar frequencies.
For higher drives, the moving stripes 
couple and the noise is more characteristic of a single
periodic signal. 
In larger systems, near depinning there could 
be larger numbers of frequencies present since there are a larger number
of stripes which can each move at different velocities.  This would broaden
the power spectrum; however, at larger drives, where the stripes couple, 
a strong narrow band noise signature should appear.

In the weak pinning system of $F_p=0.125$, $x$-direction driving produces
elastic depinning.
Just above the depinning transition, the noise signal is not periodic
as shown by the plot of $V_x$ in Fig.~18(c) for $F^{x}_{D} = 0.0375$. 
There is, however, a characteristic noise frequency, as shown by the
Lorentzian shape of $S(f)$ in Fig.~18(d).
At higher drives, the spectrum broadens and the Lorentzian peak frequency
shifts to higher frequency with increasing drive.

\begin{figure}
\includegraphics[width=3.5in]{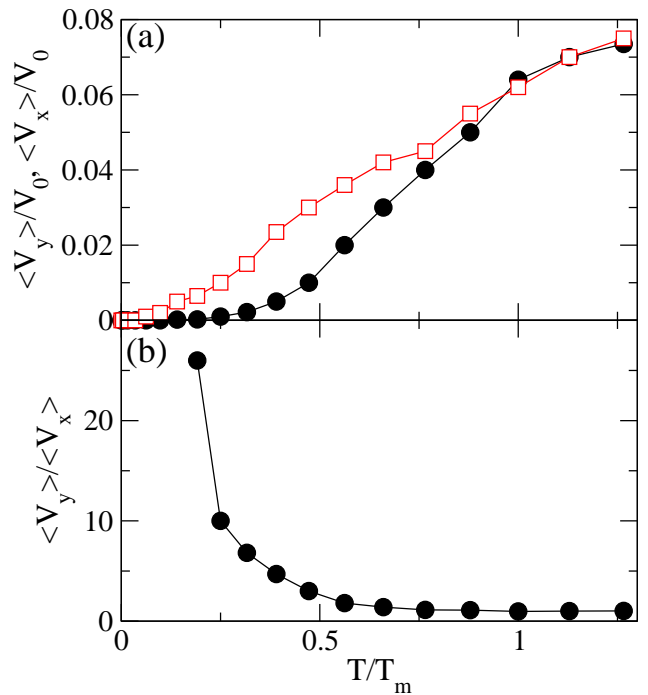}
\caption{  
(a)
$\langle V_{x}\rangle/V_{0}$ (filled circles) and 
$\langle V_{y}\rangle/V_{0}$ (open squares) vs 
$T/T_{m}$ for a system with $\rho_{p} = 0.36$, $R_{p} = 0.2$, and
$F_{p} = 0.7$ at applied drives of $F^x_{D} = 0.09$
and $F^y_D=0.09$, respectively. 
Here $T_{m}$ is the temperature at which
the stripes melt in the absence of pinning 
and $V_{0}$ is the velocity at which the particles
would move in the absence of pinning. 
Anisotropic transport occurs for $T/T_{m} < 1.0$.  
(b) Velocity anisotropy ratio $\langle V_y\rangle/\langle V_x\rangle$ vs
$T/T_m$ for the same system shows a diverging anisotropy at low temperatures.
}
\end{figure}

\begin{figure}
\includegraphics[width=3.5in]{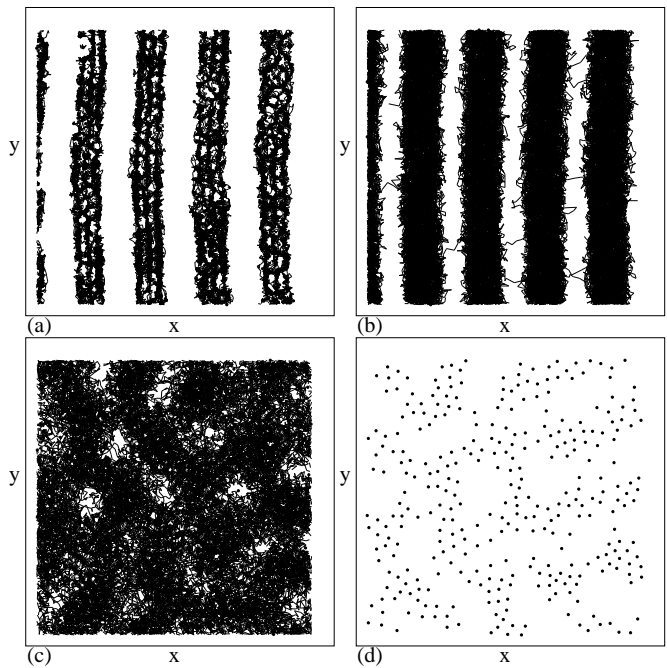}
\caption{
The particle positions (dots) and trajectories (lines) at different
temperatures for the system in Fig.~19 with $F^y_D=0.09$.
(a) At $T/T_{m} = 0.25$, 
there is creep only along the $y$-direction and no creep
in the $x$-direction.
(b) At $T/T_{m} = 0.76$, the stripe structure is still present.
There is considerable diffusion
along the stripes and a smaller amount of 
diffusion across the stripes. (c) 
At $T/T_{m} = 1.266$, the motion is no longer anisotropic, as indicated
in Fig.~19, and the diffusion is isotropic. (d) The particle
positions only from panel (c) show that the stripe structure is destroyed. 
}
\end{figure}

\section{Thermal Effects}

We next consider the  effects of thermal fluctuations. In previous work for a
stripe-forming system at the same density considered here but 
with no quenched disorder, we used diffusion and specific heat measurements to
identify a well-defined disordering temperature
$T_{m}$ above which the stripe structures were completely destroyed 
\cite{ourstuff}. Below $T_m$, there was liquid-like particle motion along
the length of the stripes, but the system behaved like a solid in the direction
perpendicular to the stripes.
This suggests that in the presence of quenched disorder, 
the stripe system might show considerable creep 
in the easy flow direction, but not in the hard direction for
a finite temperature at applied drives below the zero temperature
depinning thresholds.
To examine this, we consider
a system with $\rho_{p}= 0.36$ and $F_{p} = 0.7$. 
At drives of $F^x_{D}= 0.09$ or $F^y_D=0.09$, the system
is pinned in both directions at $T = 0.0$. 
In Fig.~19(a) we plot $\langle V_{y}\rangle/V_0$ and $\langle V_{x}\rangle/V_0$ 
versus $T/T_m$.  Here 
$V_{0}$ is the velocity at which the particles would move in the absence of 
pinning.  
For $T/T_m < 0.25$, 
there is almost no creep for $x$-direction driving but there is considerable
creep for $y$-direction driving. 
As a result, at low temperatures the anisotropy of the velocity response
diverges, as indicated by the plot of $\langle V_y\rangle/\langle V_x\rangle$
in Fig.~19(b).

In Fig.~20(a) we plot the particle positions and trajectories 
for this system with $F^y_D=0.09$ at $T/T_m=0.25$.
Here there is liquid-like motion of the particles along the stripes 
but there is no diffusion perpendicular to the
stripes. 
The stripe structure remains ordered although plastic creep is occurring.  In
the creep process, some particles along the stripe remain pinned while 
other particles move around them along the length of the stripe.
For $0.25 < T/T_{m} < 1.0$, creep occurs for both directions of drive but
there is a larger amount of creep for driving in the $y$-direction. The
anisotropy of the creep gradually diminishes as $T$ approaches $T_{m}$. 
In Fig.~20(b) we plot the particle positions
and trajectories for $F^y_D=0.09$ and $T/T_{m} = 0.76$. 
The stripe structure 
is still present but some hopping of particles from stripe to stripe 
occurs in the $x$-direction. 
For $T/T_{m} > 1.0$, the creep anisotropy vanishes.  Here
the stripe structure is completely disordered 
and there is diffusion through the entire sample 
as shown in Fig.~20(c) and Fig.~20(d) for $T/T_m=1.266$.
There is still some temporary trapping 
of particles by the pinning sites, indicated by the fact
that $\langle V_{x,y}\rangle/V_{0} < 1.0$. 
For higher temperatures, $\langle V_{x,y}\rangle/V_0$ gradually approaches $1.0$ 
as the effectiveness of the pinning is diminished.
In general we find that the creep anisotropy persists 
longer at lower pinning densities and that at higher pinning
densities the creep anisotropy disappears.

\subsection{Thermally Induced Ordering}

We find an interesting effect in which thermal noise 
induces the formation of stripe order. 
At low temperature and for sufficiently strong disorder,
the stripe structures are fragmented and destroyed.
As the temperature is increased, 
it is possible for the thermal fluctuations to wash out the 
effectiveness of the pinning 
before the melting temperature of the stripe structure is
reached.  The result is a floating ordered stripe.
A similar effect has been observed for two-dimensional vortex
\cite{Teitel} and colloid \cite{Deb} systems 
interacting with periodic and random substrates. 
In the floating solid transition found in these studies, the
vortices or colloids are pinned to the substrate at low temperatures,
while at higher temperatures they float free of the substrate 
and form a triangular lattice.  At still
higher temperatures, the lattice disorders thermally. 
When the substrate pinning is strong enough, 
the floating solid phase disappears and the
system passes directly from a pinned solid to a liquid state. 

\begin{figure}
\includegraphics[width=3.5in]{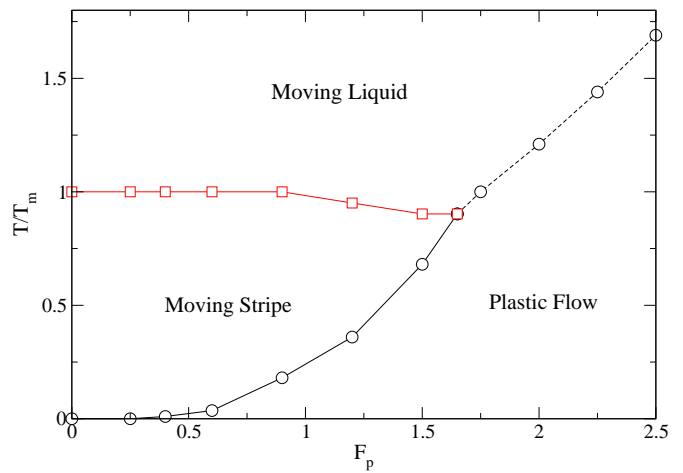}
\caption{
Phase diagram of $T/T_m$ vs $F_p$ 
for a system with $\rho_{p} = 0.38$, $R_{p} = 0.2$,
and a $y$-direction drive of $F^y_{D}= 0.22$.
Here $T_{m}$ is the melting temperature for 
the system without quenched disorder. 
The lower solid line separates the plastic flow phase 
from the moving stripe phase 
and the upper solid line separates the moving stripe phase 
from the moving liquid phase. 
The dashed line
separates the plastic flow phase from the moving liquid phase. 
The phase diagram shows that for intermediate
pinning strength, increasing the temperature can produce a transition from
a disordered plastic flow phase into an ordered moving stripe phase.
}
\end{figure}

\begin{figure}
\includegraphics[width=3.5in]{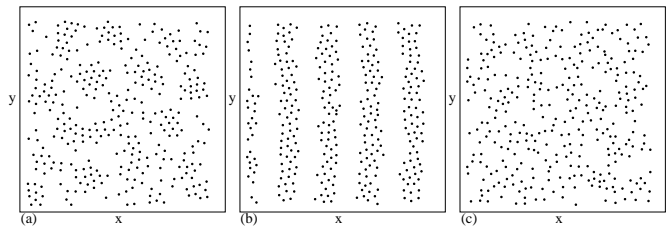}
\caption{
The particle positions (dots) for the system in Fig.~21 at 
$F_{p} = 0.9$ for different temperatures.
(a) At $T/T_{m} = 0.1$, the system is undergoing plastic flow and the particles
are disordered. 
(b) At $T/T_{m} = 0.75$, the temperature has reduced the effectiveness of the
pinning, permitting the formation of a moving stripe phase. 
(c) At $T/T_{m} = 1.3$, the
temperature is large enough to melt the stripe structure.
}
\end{figure}

To illustrate the formation of a floating stripe phase in our system in
the presence of random disorder,
in Fig.~21 we plot a phase diagram of $T/T_{m}$ versus $F_{p}$ 
for a system with $\rho_{p} = 0.38$, $R_{p} = 0.2$,
and $F^{y}_{D} = 0.22$. Here $T_{m}$ is the
melting temperature of the stripes in the absence of quenched disorder.
For low temperature and weak disorder of $F_p \leq 0.3$, the system is in a 
moving stripe phase. 
For $0.3 < F_{p} < 1.75$, at low temperatures the system is
in the strongly disordered plastic flow state illustrated in Fig.~22(a) for 
$F_{p} = 0.9$ and $T/T_{m} = 0.1$. 
As the temperature is increased, the effectiveness of the quenched
disorder is thermally destroyed and the system organizes 
into a moving stripe state aligned with the direction of drive, such
as that shown in Fig.~22(b) for $T/T_{m} = 0.75$. 
The system
melts into a liquid for $T/T_{m} > 1.0$ 
as shown in Fig.~22(c) for $T/T_{m} = 1.3$. 
The phase diagram in Fig.~21 indicates where these three 
phases occur.  
For $F_{p} > 1.6$ the system goes directly from
a partially pinned plastic flow phase to 
an unpinned 
disordered liquid phase, as indicated by the dashed line.        
This result suggests that in systems with strong disorder, it is possible that
the anisotropy may be weak at low temperatures but could increase for
intermediate temperatures when the floating moving stripe structure forms.

The ability of the system to form a floating stripe depends on the 
length scale of the quenched disorder as well as the disorder strength.
If the disorder is composed of 
small well localized pins, as in our model, a 
floating solid phase is possible. 
In contrast, for long range quenched disorder the thermal noise will
not be effective in washing out the 
pinning and a floating solid phase will not occur.
The presence of a floating solid phase makes it impossible to observe
a thermally induced peak effect phenomena of the type
found in superconducting vortex systems.
In this peak effect, a thermally melted vortex lattice is softer and
is able to better couple to the quenched disorder, increasing the
depinning force \cite{Ling}.
If the thermal fluctuations destroy the effectiveness of the
quenched disorder below the temperature at which the disorder-free equilibrium
particle structure melts, 
a thermally induced peak effect can not occur. 
It is possible that adding
long range correlations to the pinning 
would permit the appearance of a thermally induced peak effect 
in the stripe forming system.  

\section{Summary}

We have examined the anisotropic dynamics of 
oriented stripes in a system with competing interactions.
We focus on the regime where stripe structures form in equilibrium and in
the absence of quenched disorder. 
After adding a random substrate, we drive the system parallel or perpendicular
to the original orientation of the stripes.
We find 
anisotropic depinning thresholds and nonlinear
velocity force curves. 

Under the sudden application of an external drive,
the system settles into a steady state flow after a transient time that
is determined by the structure of the steady state flow, such as a 
plastically flowing state or a moving ordered state. 
The transient time passes through peaks at the transitions between
different dynamical states, such as from pinned to filamentary
flow or from strongly fluctuating plastic flow to ordered flow. 
In addition to the stripe system, this type of transient measurement after a
sudden application of a drive could also be used in other driven systems
such as superconducting vortices with quenched disorder, friction, and
sliding charge density waves.

We observe different types of plastic flow which are determined by the
direction of the drive relative to the stripe orientation.
For driving parallel to the stripes, there is
a phase in which the stripes 
remain ordered but are decoupled and can slide past one another. 
For stronger quenched disorder, plastic flow can occur within
individual stripes while the overall stripe structure remains intact. 
In this case, the flow is filamentary and involves only a portion of the
particles within the stripe.
For stronger or denser quenched disorder, 
the stripes break apart and we find a strongly fluctuating plastic flow
phase in which the transport properties are isotropic.
For driving perpendicular to the stripe orientation, in addition to
plastic flow phases there can be elastic depinning of the 
stripes perpendicular to the drive for sufficiently weak disorder.     

As a function of disorder strength we find a sharp order to disorder 
transition in which the state above depinning changes from an ordered
moving stripe structure to a plastic flow regime which tears apart
the stripes.
This order-disorder transition is accompanied by a
sharp increase in the depinning threshold 
which is similar to the peak effect phenomenon 
observed near order-disorder transitions
for vortex matter in type-II superconductors. 
In the stripe system the order-disorder transition occurs
at different disorder strengths for the two different driving directions, 
producing regimes of enhanced anisotropy in which 
the system depins plastically in one direction but elastically in the
other.

In the plastic flow regime near depinning, we observe a series 
of velocity jumps and transitions which correspond to transitions between
filamentary flow states associated with narrow band time-of-flight velocity 
noise signatures
and strongly fluctuating plastic flow states exhibiting broad band noise
signatures.
These transitions are very similar to recent experimental 
observations in this class of system. 

The anisotropic transport can be enhanced by thermal fluctuations.
Thermal disorder induces an anisotropic melting of the stripes, with a 
lower temperature stripe liquid in which particles can move easily along
the length of the stripe but remain confined perpendicular to the stripe,
and a higher temperature isotropic liquid in which the stripe structure is
destroyed.
For intermediate quenched disorder strength, 
the stripe structure is disordered at low temperatures but can
undergo a thermally induced stripe ordering into a floating stripe 
phase when the thermal fluctuations reduce the effectiveness of
the quenched disorder.

We expect that these results should be generic to any type of 
stripe forming system driven over quenched disorder. Particular
systems where the pulse measurements and transient times 
could be analyzed include two-dimensional electron gasses or the
recently studied type-1.5 superconductors in which the vortices interact
via competing repulsive and attractive interactions.
Other relevant systems include stripe or labyrinth patterns 
in soft matter systems driven with electric or
magnetic fields over a rough surface or through obstacle arrays.      

This work was carried out under the auspices of the 
NNSA of the 
U.S. DoE
at 
LANL
under Contract No.
DE-AC52-06NA25396.


\begin{thebibliography}{99}

\bibitem{Andleman}
M.~Seul and D.~Andelman, Science {\bf 267}, 476 (1995). 

\bibitem{Boyer}
D.~Boyer and J.~Vi\~nals, Phys.~Rev.~E {\bf 64}, 050101(R) (2001). 

\bibitem{Singer}
A.D.~Stoycheva and S.J.~Singer, Phys.~Rev.~Lett.~{\bf 84}, 4657 (2000).
A.D.~Stoycheva and S.J.~Singer, Phys.~Rev.~E {\bf 65}, 036706 (2002).

\bibitem{Pell}
G.~Malescio and G.~Pellicane, Nature Mater. {\bf 2}, 97 (2003).  

\bibitem{Glaser}
M.A.~Glaser, G.M. Grason, R.D. Kamien, A. Kosmrlj, C.D. Santangelo,
and P. Ziherl, EPL {\bf 78}, 46004 (2007).  

\bibitem{Colloid}
W.M.~Gelbart, R.P.~Sear, J.R.~Heath, and S.~Chaney, 
Farady Discuss. {\bf 112}, 299 (1999);
M.~Klokkenburg, R.P.A.~Dullens, W.K.~Kegel, B.H.~Ern{\' e}, and A.P.~Philipse,
Phys.~Rev.~Lett.~{\bf 96}, 037203 (2006).

\bibitem{S}
E.A. Jagla, Phys.~Rev.~E {\bf 58}, 1478 (1998);   
J. Fornleitner and G. Kahl, EPL {\bf 82}, 18001 (2008);
H. Shin, G.M.~Grason, and C.D.~Santangelo, Soft Matter {\bf 5}, 3629 (2009). 

\bibitem{Colloid2}
P.J.~Camp, Phys.~Rev.~E {\bf 68}, 061506 (2003);
P.J.~Camp, Phys. Rev.~E {\bf 71}, 031507 (2005). 

\bibitem{Zapperi}
A.~Imperio, L.~Reatto and S.~Zapperi,
Phys.~Rev.~E {\bf 78}, 021402 (2008). 

\bibitem{Peeters}
K.~Nelissen, B.~Partoens, and F.M.~Peeters,
Phys.~Rev.~E {\bf 71}, 066204 (2005);
F.F.~Munarin, K.~Nelissen, W.P.~Ferreira, G.A.~Farias, and 
F.M.~Peeters, {\it ibid.} {\bf 77}, 031608 (2008); 
Y.H.~Liu, L.Y.~Chew, and M.Y.~Yu,
Phys.~Rev.~E {\bf 78}, 066405 (2008).  

\bibitem{J}
J.~Zaanen and O.Gunnarsson, Phys.~Rev.~B {\bf 40}, 7391 (1989).

\bibitem{Stripe}
J.M.~Tranquada, B.J.~Sternlieb, J.D.~Axe, Y.~Nakamura, and S.~Uchida,
Nature(London) {\bf 375}, 561 (1995).

\bibitem{SA}
S.A.~Kivelson, E.~Fradkin, and V.J.~Emery,
Nature {\bf 393}, 550 (1998).

\bibitem{Bishop}
B.P.~Stojkovic, Z.G.~Yu, A.L.~Chernyshev, A.R.~Bishop, A.H.~Castro Neto,
and N. Gr{\o}nbech-Jensen, Phys.~Rev.~B {\bf 62}, 4353 (2000). 

\bibitem{Kabanov}
T.~Mertelj, V.V.~Kabanov, and D.~Mihailovic,
Phys.~Rev.~Lett.~{\bf 94}, 147003 (2005). 

\bibitem{Stroud}
D.~Valdez-Balderas, and D.~Stroud,
Phys.~Rev.~B {\bf 72}, 214501 (2005).

\bibitem{Kivelson}
M.M.~Fogler, A.A.~Koulakov, and B.I.~Shklovskii, Phys.~Rev.~B {\bf 54},
1853 (1996);
R.~Moessner and J.T.~Chalker, Phys.~Rev.~B {\bf 54}, 5006 (1996); 
E.~Fradkin and S.A.~Kivelson, Phys.~Rev.~B {\bf 59}, 8065 (1999).

\bibitem{Bishop2}
C.J. Olson Reichhardt, C.~Reichhardt, and A.R.~Bishop,
Phys.~Rev.~Lett.~{\bf 92}, 016801 (2004).

\bibitem{Bishop3}
C.J.~Olson Reichhardt, C.~Reichhardt, I.~Martin, and
A.R.~Bishop, Physica D {\bf 193}, 303 (2004).  

\bibitem{Martin}
C.~Reichhardt, C.J. Olson Reichhardt, I.~Martin, and A.R.~Bishop,
Phys.~Rev.~Lett.~{\bf 90}, 026401 (2003).

\bibitem{Olson}
C. Reichhardt, C.J. Olson, I. Martin and A.R. Bishop,
Europhys. Lett. {\bf 61}, 221 (2003).

\bibitem{Lilly}
M.P.~Lilly, K.B.~Cooper, J.P.~Eisenstein, L.N.~Pfeiffer, and K.W.~West,
Phys.~Rev.~Lett.~{\bf 82}, 394 (1999).

\bibitem{Du}
R.R.~Du, D.C.~Tsui, H.L.~Stormer, L.N.~Pfeiffer, K.W.~Baldwin, and K.W.~West,
Sol.~St.~Commun. {\bf 109}, 389 (1999).

\bibitem{JE}
M.P.~Lilly, K.B.~Cooper, J.P.~Eisenstein, L.N.~Pfeiffer, and K.W.~West,
Phys.~Rev.~Lett.~{\bf 83}, 824 (1999). 

\bibitem{M}
K.B.~Cooper, J.P.~Eisenstein, L.N.~Pfeiffer, and K.W.~West, 
Phys.~Rev.~Lett. {\bf 92}, 026806 (2004).  

\bibitem{Co}
K.B.~Cooper, M.P.~Liily, J.P.~Eisenstein, 
T.~Jungwirth, L.N.~Pfeiffer, and K.W.~West,
Sol.~St.~Commun.~{\bf 119}, 89 (2001).

\bibitem{Gores}
J.~G{\" o}res, G.~Gamez, J.H.~Smet, L.~Pfeiffer, K.~West, A.~Yacoby, V.~Umansky,
and K. von Klitzing, Phys.~Rev.~Lett.~{\bf 99}, 246402 (2007).

\bibitem{Cheong}
Y.~Horibe, C.H.~Chen, S.-W.~Cheong, and S.~Mori,
Europhys. Lett. {\bf 70}, 383 (2005).

\bibitem{Gabor}
S.P.~Koduvayur, Y.~Lyanda-Geller, S.~Khlebnikov, G.~Csathy, M.J.~Manfra,
L.N.~Pfeiffer, K.W.~West, and L.P.~Rokhinson,
 arXiv:1005.3327. 

\bibitem{Cooper}
K.B.~Cooper, M.P.~Lilly, J.P.~Eisenstein, L.N.~Pfeiffer, and K.W.~West,
Phys.~Rev.~B {\bf 60}, R11285 (1999).

\bibitem{N}
K.B.~Cooper, J.P.~Eisenstein, L.N.~Pfeiffer, and K.W.~West,
Phys.~Rev.~Lett.~{\bf 90}, 226803 (2003). 

\bibitem{Lewis}
G.~Sambandamurthy, R.M.~Lewis, H.~Zhu, Y.P.~Chen, L.W.~Engel,
D.C.~Tsui, L.N.~Pfeiffer, and K.W.~West, 
Phys.~Rev.~Lett.~{\bf 100}, 256801 (2008).

\bibitem{Zhu}
H.~Zhu, G.~Sambandamurthy, L.W.~Engel, D.C.~Tsui, L.N.~Pfeiffer,
and K.W.~West, Phys.~Rev.~Lett.~{\bf 102}, 136804 (2009).

\bibitem{Mosch}
V.~Moshchalkov, M.~Menghini, T.~Nishio, Q.H.~Chen, A.V.~Silhanek, V.H.~Dao,
L.F.~Chibotaru, N.D.~Zhigadlo, and J.~Karpinski, 
Phys.~Rev.~Lett.~{\bf 102}, 117001 (2009);
E.H.~Brandt and S.-P.~Zhou, Physics {\bf 2}, 22 (2009); 
T.~Nishio, V.H.~Dao, Q.~Chen, L.F.~Chibotaru, K.~Kadowaki, and
V.V.~Moshchalkov, Phys.~Rev.~B {\bf 81}, 020506(R) (2010). 

\bibitem{Mosch2}
E. Babaev and M. Speight, Phys.~Rev.~B {\bf 72}, 180502(R) (2005).

\bibitem{Ling2}
C.~Reichhardt and C.J. Olson, Phys.~Rev.~Lett.~{\bf 89},
078301 (2002); 
A.~Pertsinidis and X.S.~Ling, {\it ibid.} {\bf 100}, 028303 (2008). 

\bibitem{Hu}
J.~Tekic, O.M.~Braun, and B.~Hu, Phys.~Rev.~E {\bf 71}, 026104 (2005). 

\bibitem{Peak}
A.B.~Pippard, Philos.~Mag.~{\bf 19}, 217 (1969).

\bibitem{Higgins}
S.~Bhattacharya and M.J.~Higgins, Phys.~Rev.~Lett.~{\bf 70}, 2617 (1993).

\bibitem{Ling}
X.S.~Ling, S.R.~Park, B.A.~McClain, S.M.~Choi, 
D.C.~Dender and J.W.~Lynn,
Phys.~Rev.~Lett.~{\bf 86}, 712 (2001).

\bibitem{P}
S.~Mohan, J.~Sinha, S.S.~Banerjee and Y.~Myasoedov,
Phys.~Rev.~Lett.~{\bf 98}, 027003 (2007).

\bibitem{Koshelev}
A.E.~Koshelev and V.M.~Vinokur,
Phys.~Rev.~Lett.~{\bf 73}, 3580 (1994).

\bibitem{Balents}
T.~Giamarchi and P. Le Doussal, Phys.~Rev.~Lett.~{\bf 76}, 3408 (1996);
S.~Scheidl and V.M.~Vinokur, Phys.~Rev.~E {\bf 57}, 2574 (1998); 
L.~Balents, M.C.~Marchetti, and L.~Radzihovsky, Phys.~Rev.~B {\bf 57}, 7705 (1998).

\bibitem{Zimanyi}
K.~Moon, R.T.~Scalettar, and G.T.~Zim{\' a}nyi, 
Phys.~Rev.~Lett.~{\bf 77}, 2778 (1996);  
S.~Ryu, M.~Hellerqvist, S.~Doniach, A.~Kapitulnik, and
D.~Stroud, Phys.~Rev.~Lett.~{\bf 77}, 5114 (1996);
A.B.~Kolton, D.~Dom{\' \i}nguez, and N.~Gr{\o}nbech-Jensen, {\it ibid.}
{\bf 83}, 3061 (1999).

\bibitem{Olson2}
C.J.~Olson, C.~Reichhardt, and F.~Nori,
Phys.~Rev.~Lett.~{\bf 81}, 3757 (1998).

\bibitem{Marchetti}
M.C.~Faleski, M.C.~Marchetti, and A.A.~Middleton, 
Phys.~Rev.~B {\bf 54}, 12427 (1996).

\bibitem{Pardo}
F.~Pardo, F.~de la Cruz, P.L.~Gammel, E.~Bucher, and D.J.~Bishop,
 Nature (London) {\bf 396}, 348 (1998).

\bibitem{Brass}
H.J.~Jensen, A.~Brass, Y.~Brechet, and A.J.~Berlinsky, 
Phys.~Rev.~B {\bf 38}, 9235 (1988);
P.~Moretti and M.~Carmen Miguel, Phys.~Rev.~B {\bf 79}, 104505 (2009).

\bibitem{Du2}
R.~Danneau, A.~Ayari, D.~Rideau, H.~Requardt, J.E.~Lorenzo, L.~Ortega,
P.~Monceau, R.~Currat, and G.~Gr{\" u}bel, 
Phys.~Rev.~Lett.~{\bf 89}, 106404 (2002); 
C-H.~Du, C.-Y.~Lo, H.-H.~Lin and S.L.~Chang,
J.~Appl.~Phys. {\bf 101}, 104915 (2007).

\bibitem{Jensen}
C.~Reichhardt, C.J.~Olson, N.~Gr{\o}nbech-Jensen, and F.~Nori,
Phys.~Rev.~Lett.~{\bf 86}, 4354 (2001).

\bibitem{Lekner}
J. Lekner, Physica A {\bf 176}, 485 (1991); 
N. Gr{\o}nbech-Jensen, Int. J. Mod. Phys. C {\bf 8}, 1287 (1997);
M. Mazars, Mol. Phys. {\bf 103}, 1241 (2005).

\bibitem{ourstuff}
C.J.~Olson Reichhardt, C.~Reichhardt, and A.R.~Bishop,
Phys.~Rev.~E {\bf 82}, 041502 (2010).

\bibitem{Corte}
L. Cort{\' e}, P.M.~Chaikin, J.P.~Gollub, and D.J.~Pine, 
Nature Phys.~{\bf 4}, 420 (2008).

\bibitem{R}
C. Reichhardt and C.J. Olson Reichhardt, 
Phys.~Rev.~Lett.~{\bf 103}, 168301 (2009).

\bibitem{Andrei}
Z.L.~Xiao, E.Y.~Andrei, and M.J.~Higgins, Phys.~Rev.~Lett.~{\bf 83}, 
1664 (1999).

\bibitem{Scalettar}
C.J.~Olson, C. Reichhardt, R.T.~Scalettar, G.T.~Zim{\' a}nyi, 
and N. Gr{\o}nbech-Jensen,
Phys.~Rev.~B {\bf 67}, 184523 (2003). 

\bibitem{Krauth}
O. Duemmer and W.~Krauth, Phys.~Rev.~E {\bf 71}, 061601 (2005). 

\bibitem{Vinokur}
V.M.~Vinokur and T.~Nattermann, Phys.~Rev.~Lett.~{\bf 79}, 3471 (1997). 

\bibitem{M2}
M.C.~Marchetti, A.A.~Middleton, and T.~Prellberg,
Phys.~Rev.~Lett.~{\bf 85}, 1104 (2000).  

\bibitem{Saunders}
K.~Saunders, J.M.~Schwarz, M.C.~Marchetti, and A.A.~Middleton,
Phys.~Rev.~B {\bf 70}, 024205 (2004).

\bibitem{MC}
P.~Le Doussal, M.C.~Marchetti and K.J.~Wiese,
Phys.~Rev.~B {\bf 78}, 224201 (2008).

\bibitem{Brandt}
E.H.~Brandt, J. Low Temp. Phys. {\bf 53}, 41 (1983).

\bibitem{H}
M.C.~Hellerqvist, D.~Ephron, W.R.~White, M.R.~Beasley, and A.~Kapitulnik,
Phys.~Rev.~Lett.~{\bf 76}, 4022 (1996).

\bibitem{Moh}
S.~Mohan, J.~Sinha, S.S.~Banerjee, A.K.~Sood, S.~Ramakrishnan, and A.K.~Grover,
Phys.~Rev.~Lett.~{\bf 103}, 167001 (2009).

\bibitem{SB}
M.-C.~Cha and H.A.~Fertig, Phys.~Rev.~Lett.~{\bf 80}, 3851 (1998);
C.J.~Olson, C.~Reichhardt, and S.~Bhattacharya, 
Phys.~Rev.~B {\bf 64}, 024518 (2001); 
M.~Chandran, R.T.~Scalettar, and G.T.~Zim{\' a}nyi, 
Phys. Rev.~B {\bf 67}, 052507 (2003).  

\bibitem{Menon}
A.~Sengupta, S.~Sengupta, and G.I.~Menon, 
Phys.~Rev.~B {\bf 81}, 144521 (2010).

\bibitem{GT}
G.A.~Cs{\' a}thy, D.C.~Tsui, L.N.~Pfeiffer, and K.W.~West,
Phys.~Rev.~Lett.~{\bf 98}, 066805 (2007).

\bibitem{Dom}
N. Gr{\o}nbech-Jensen, A.R.~Bishop, and D. Dom{\' \i}nguez, 
Phys.~Rev.~Lett.~{\bf 76},
2985 (1996).

\bibitem{Am}
L.~Ammor, A.~Ruyter, V.A.~Shaidiuk, N.H.~Hong, and D.~Plessis,
Phys.~Rev.~B {\bf 81}, 094521 (2010).

\bibitem{Soret}
E.~Olive and J.C.~Soret, Phys.~Rev.~B {\bf 77}, 144514 (2008).

\bibitem{Teitel}
M.~Franz and S.~Teitel, Phys.~Rev.~B {\bf 51}, 6551 (1995);
S.A.~Hattel and J.M.~Wheatley, Phys.~Rev.~B {\bf 51}, 11951 (1995);
C.~Reichhardt, C.J.~Olson, R.T.~Scalettar, 
and G.T.~Zim{\' a}nyi, Phys.~Rev.~B {\bf 64}, 144509 (2001).  

\bibitem{Deb}
D.~Deb and H.H.~van Gr{\"u}nberg, J. Phys.: Condens. Matter 
{\bf 20}, 245104 (2008);
S.~Herrea-Velarde and H.H.~von Gr{\"u}nberg, Soft Matter {\bf 5}, 391 (2009). 

\end{thebibliography}
\end{document}